\documentclass{iopjournal}
\usepackage{amsmath}
\usepackage{amsfonts}
\usepackage{graphicx}
\usepackage{float}
\usepackage{color}
\usepackage{doi}
\usepackage{bm}
\usepackage{array}
\usepackage{mathtools}
\usepackage[T1]{fontenc}
\usepackage{bigints}
\usepackage{url}
\usepackage{notoccite}
\usepackage{cite}
\usepackage[symbol]{footmisc}
\usepackage{bm}
\usepackage[british]{babel}

\begin{document}
	
	\title{A quantum optical concept of attosecond pulses: the attoquants}

	\author{Varró Sándor$^{1}$\orcid{0000-0002-7246-7369},Gombkötő Ákos$^{2}$\orcid{0000-0002-4054-9060}}
	
	\affil{$^1$ELI ALPS, The Extreme Light Infrastructure ERIC, Wolfgang Sandner u. 3., 6728 Szeged, Hungary}
	
	\affil{$^2$HUN-REN Wigner Research Centre for Physics, Konkoly-Thege M. \'{u}t 29-33, H-1121 Budapest, Hungary}
	\email{varro.sandor@eli-alps.hu, gombkoto.akos@wigner.hun-ren.hu}

	\begin{abstract}
		High-harmonic generation (HHG) is conventionally understood to produce attosecond pulse trains only when the emitted harmonics are mutually phase-locked, in accordance with the classical theory of mode-locking. In a quantum optical framework, pulse synthesis would also require a regular phase relation between the components, when the harmonic field is in a separable multimode quantum state. This requirement, however, can be circumvented for harmonic combs exhibiting intermodal entanglement. We introduce a novel family of multimode quantum states, termed attoquants: states associated with highly structured pulse trains that are inherently insensitive to relative phases. As a concrete example, we analyse the coherent permanent state, constructed as a completely symmetric superposition of products of coherent states over all permutations of a fixed parameter set. It is shown analytically that its electric-field mean value is a locked Fourier superposition of the driving-field harmonics, yielding an attosecond pulse train without requiring regular phase relations between the high-harmonic components. We further calculate the photon statistics, the Wigner function, and the logarithmic negativity of these states, in order to show their genuinely nonclassical character, and the degree of entanglement. Finally, we offer a phenomenological interpretation for the generation of such entangled states during HHG, as a consequence of multi-atom effects. Since the coherence volume of the driving field greatly exceeds that of its harmonics, a collectively-driven cluster of atoms can radiate the harmonics as spatially separated, but classically indistinguishable sources. This can generate the permutation-symmetric entanglement structure of an attoquant.
	\end{abstract}
	
	\section{Introduction}\label{sec:intro}
	
	High-harmonic generation (HHG) is a cornerstone of attophysics. During HHG electromagnetic pulses drive nonlinear electron dynamics, producing radiation spanning a broad spectral range. 
	Attosecond light pulses are synthesised from the high-order harmonics generated during this light-matter interaction, involving material targets such as rare-gas jets~\cite{FMLA88}. 
	In the simplest picture, the different harmonics are emitted during the recombination of highly excited electrons of different, but ``synchronised'' atoms. Thus, one may assume a regular phase relation between the harmonics, whose classical Fourier-synthesis yields attosecond pulse trains~\cite{FT92}. 
	As is known, the phase parameters are of utmost
	importance for the production of attosecond pulses. Already at the early stage of research on high-harmonic generation, the importance of the proper phase-matching of the generated radiation has been stressed~\cite{PhysRev.185.57,10.1063/1.339157,10.1007/3-540-51430-9_6,Balcou1993,antoine1997phase,PSalieres_1996}. Within the classical framework, mode-locking is indispensable; any deviation in the relative phases of the modes impairs the Fourier synthesis required to form the pulse train. Since, according to theory, the dipole-phase is an approximately linear function of intensity, itself a function of time and space, the prospects of maintaining a constant phase difference between the harmonics were a priori rather poor \cite{PierreAgostini_2004}. 
	Therefore, the existence of phase-locking between high-order harmonics was initially questioned by many researchers~\cite{varro2024multiphoton}. 
	It was subsequently demonstrated experimentally~\cite{doi:10.1126/science.1059413}. Theory and experiment agree qualitatively, once the propagation of the harmonics through the medium is taken into account~\cite{PhysRevLett.77.1234}, provided the medium's dimensions are small compared with the beam's Rayleigh range. Then, phase-matching \cite{PhysRevLett.66.2200} can be understood as the interplay between the geometrical phase shift of the driving field (the Gouy phase) and the variation of the dipole-phase that follows its intensity distribution \cite{PhysRevLett.74.3776}. 
	The mode-locking of the harmonics understood in this sense has since become a well-established feature in the field of attophysics~\cite{PhysRevLett.94.033001}.

	In order to illustrate the situation concerning the experimental research of multiphoton processes, let us quote Lompré, L’Huillier and Mainfray \cite{10.1007/3-540-51430-9_6} from one of the earliest papers on high-harmonic generation: \textit{"Ion and electron experiments are performed at low pressure, typically 10$^{-5}$ Torr, about $3\times10^{11}$ atoms/cm$^3$, in order to minimize collective effects. Due to the poor collection efficiency of the photon detection, photon experiments require to be performed at a much higher atomic density, typically 10$^{17}$ to 10$^{18}$ atoms/cm$^3$, about 10 to 20 Torr. At such a high pressure, the laser light does not interact anymore with isolated atoms, but with a collective assembly of atoms. The emitted radiation may be due to fluorescence from excited states of atoms or ions either directly by the laser or by various plasma recombination mechanisms. The additional processes due to the high pressure turn out to be dominant effects. Provided the medium has the required pressure and geometrical properties for proper phase-matching of the generated radiation, one may expect production of high-order harmonics of the laser field."}
	Currently, the dominant view in the attophysics community is that a combination of microscopic single-atom responses and macroscopic Maxwell's equations provides a sufficient model for most purposes. Such an approach, however, may overlook multi-atom effects~\cite{PhysRevA.41.6571,Farkas1993}, the account of which may be necessary for a comprehensive understanding of the resulting pulse-structures.
	
	In the present paper, we discuss attosecond pulse synthesis within the framework of quantum optics. Since the aforementioned recombination is in fact a spontaneous emission process~\cite{ehlotzky1992harmonic,varro1993generation,varro1993new,photonics8070269}, understanding the locking of the different harmonics is far from trivial from a quantum optical perspective. Likely, the process of collective emission plays a key role in the seemingly ``synchronised'' atomic emissions. 
	As we will show, some quantum states --that can be qualitatively connected to such collective emissions-- possess an inherently robust pulse structure: even a random distribution of the phases results in a locking of the temporal Fourier components. 
	We coin the term \emph{attoquants} (or attoquant states) to label quantum states corresponding to such a phenomenon. A realisation of attoquants is the entangled, completely symmetric sum of coherent state products, comprising the focus of the current work.
	The idea of attoquants was introduced by one of the authors~\cite{Varrolphystalk,VarroATTO2019} earlier. In the cited presentations, one of the most important features of the symmetric entangled coherent states, namely the “automatic locking” of the mean-field pulses, was demonstrated using the same quantum optical framework that we develop fully here.
	To date, these states have not been discussed in the peer-reviewed literature. We believe that our analysis provides a useful contribution towards a rigorous quantum optical foundation for the notion of a nonclassical pulse train.
	We will show how, at a phenomenological level, similar states can plausibly be associated with multi-atom effects influencing HHG in gaseous media.
	
	It has become increasingly evident that the HHG process can both be modified by nonclassical light and be used to generate radiation with genuinely nonclassical properties. 
	The quantum theoretical treatment of HHG dates back to 1981~\cite{JV99}. It was followed by an investigation of Keldysh-type approaches~\cite{ehlotzky1992harmonic} in 1992, and calculations in the multiphoton Kramers-Henneberger frames~\cite{varro1993new,varro1993generation} in 1993. A quantised three-step model was devised in 1999~\cite{KMOV99} and scattering-theoretical treatments have been done in 2000~\cite{GSE2000,GLFG2000}. Research resumed after 2011~\cite{PhysRevLett.106.023001,PhysRevA.89.063827,YZSSG15}, and the field has expanded rapidly since 2020.
	Nonclassical properties induced in the harmonic radiation or in the driving field during HHG, together with the effects due to a nonclassical driving field, or due to correlations within the target material, have since been documented extensively~\cite{JV99,VS2008,VS10,GC16,GO16,TNKIGIT17,TKDF19,G20,AG19,GFV21Q,GAthesis,LCPRSKT21,RSPLTPLC21,photonics8070269,PhysRevA.105.033714,PhysRevLett.128.047401,SPRJLTL21,STPH22,PhysRevResearch.5.043138,Gombk2023,PhysRevA.108.053119,Pizzi2023,PhysRevA.110.023115,PRXQuantum.5.010328,PhysRevB.109.125110,Hansen_2024,PhysRevA.109.012223,PhysRevA.109.033110,PhysRevA.109.063103,PhysRevB.109.035203,PhysRevA.110.063704,PhysRevA.109.033706,PhysRevA.110.063118,Stammer_2024,PhysRevLett.132.143603,PRXQuantum.5.040319,PhysRevResearch.6.033010,3gjp-f7br,PhysRevLett.134.013601,11vz-9gcz,6r6n-pxfp,PhysRevX.15.011023,gombkoto2026intermodalentanglementquantumoptical,Varro2022squeezing}. Nevertheless, the pursuit of a fully quantised treatment of attophysics in general, and of HHG in particular, remains an ongoing endeavour \cite{RDJPLLMSP26,Dahl2026nat,Dubois:2026lzs,hack2026wavepacketmotionquantized}. Experimental set-ups and results are detailed in~\cite{PhysRevA.89.063827,TKDF19,photonics8060192,CruzRodrguez2024QuantumPI,Lemieux2025,Rasputnyi2024,PRXQuantum.5.040319,Lamprou_2025,Nayak2025,6r6n-pxfp}, while reviews and summaries are available in~\cite{Gombkt2021AMK,8thAttoSci,SHI2025100577,Stammer_Lewenstein_2023,doi:10.1142/9789811279560_0007,Rivera+2025+1837+1855,FennelT21,Tzallas2023,Tzallasa2023,KDPCorkum23,Ciappina2025,photonics8070263,photonics8070269,43047dc146e64dd1980cdee6d66fe5b8,PhysRevA.110.043119,PhysRevA.111.013113,PhysRevA.110.043115,11vz-9gcz,Lamprou_2025}. 
	
	Whilst the presence of nonclassical attributes in harmonic radiation is now well established at the level of individual photonic modes and mode pairs, comparatively little attention has been paid to the collective, many-mode properties of the resulting nonclassical pulse trains. The Hilbert space of the electromagnetic field is vastly larger than its classical counterpart~\cite{FCGGAA10} wherein a state is uniquely characterised by a sequence of complex parameters $\{\alpha_n\}$.
	The space of multimode quantum radiation remains largely uncharted, and a systematic exploration is far from complete. Recent interest in bright squeezed-vacuum pulses and other non-coherent-state driving fields~\cite{GTBK22,TBGK22,Rasputnyi2024,Stammer2024,PhysRevResearch.6.L032033,PhysRevResearch.6.033079,4807f9d595674c25a8147aeb297df08f,PhysRevA.111.063105,4hdl-bdwj,PhysRevA.111.043111} further motivates a re-examination of the fundamental characteristics of short pulses.
	Thus, a quantum theoretical analysis of the temporal structure of attosecond pulses remains an open task in the field of strong-field quantum optics. In a sense, this brings the discussion back to the question posed at the very outset of this Introduction: the origin of the regular phase relation on which the classical picture of the pulse train depends. As shown in this work, for suitable quantum states this dependence need not hold at all.
	
	
	The paper is organised as follows: Section~\ref{sec:model} introduces the simplified model we use to analyse the attosecond pulse trains. In section~\ref{sec:permanent}, we define and characterise the symmetric entangled coherent states. We interpret these states as attoquants in section~\ref{attoquant}. A discussion of the possible physical origins of attoquants follows in section~\ref{sec:dicsussion}. A summary of our results is offered in section~\ref{sec:conclusions}.

	\section{Simple quantum optical models of attosecond pulses}\label{sec:model}
	In this section, we first present the model of the quantised harmonic radiation field used throughout this work, and revisit the concept of pulses within this quantum optical framework.
	It is well known that a product of coherent states gives the most straightforward connection to the classical description of light pulses~\cite{PhysRev.130.2529}. Mode-locking, a key concept in the classical theory of attosecond pulse generation, can be embedded into the quantised framework without difficulty, using such states.
	However, pulses with no classical counterpart also exist, and their structural dependence on mode-locking is less obvious. We give a brief, general treatment of pulse synthesis, and demonstrate that the classical requirement of mode-locking among the harmonic modes need not persist for an arbitrary quantum state: it may be relaxed if the harmonic modes have intermodal entanglement. 
	This section is a preparatory introduction to section~\ref{sec:permanent} and section~\ref{attoquant}, where we define a special quantum state, and show that it corresponds to a pulse train with phase-insensitive structure, respectively.
	
	\subsection{Representation of the harmonic field}\label{representationharmonic}
	In this work, we model the generated harmonic radiation field as a finite collection of quantised electromagnetic modes with discrete frequencies $(2q+1)\omega$. Here $q \in \{n_0, \dots, N\}$ is an integer, that is, we consider only odd-order harmonics of the driving field of fundamental frequency $\omega$.
	The indices $n_0$ and $N$ denote the order of the lowest and highest harmonics considered, $(2n_0+1)$ and $(2N+1)$, respectively. The spatial mode structure is chosen to be that of a plane wave. All modes are taken to share an identical linear polarisation $\bm{e}^{(\mu)}$ and wave-vector direction. 
	As such, these quantum states span a finite tensor-product Fock space $\mathcal F = \bigotimes_{n=1}^{N-n_0+1}\mathcal H_n$.
	We note that even-order harmonics, Mollow sidebands, and harmonics of multi-chromatic driving fields may be incorporated into this formalism in a straightforward manner. 
	
	Throughout the paper, we neglect the classical dispersion of the high harmonics. As is known, the frequency-dependence of the index of refraction influences the phase-matching condition and propagation of the harmonics. However, for high enough frequencies, the index of refraction may be approximated with the Drude formula,  which, for the $j$th harmonic frequency, can be written as $n(j\omega) =\sqrt{1- \omega^2_p/(j\omega)^2}$, with $\omega_p$ being the plasma frequency~\cite{PhysRevA.53.1725}. 
	Since $\hbar\omega_p$ is of the order of $1~\mathrm{eV}$ at densities of $10^{20}~\mathrm{cm}^{-3}$, the deviation from the vacuum value remains small at lower, realistic densities (say $10^{18}~\mathrm{cm}^{-3}$) typical of gas-jet setups. We note that dispersion is further reduced by the blue-shift of the generated harmonics~\cite{photonics8070269}. This justifies ignoring frequency dependence as a first approximation, without affecting our conclusions.
	
	Having established these approximations, the general quantum state of the multimode harmonic field can be expressed as
	\begin{equation}\label{genquantumstate}
		|\Psi(t)\rangle
		= \sum_{\mathcal{F}}
		C_{q_1,\ldots,q_{M}}(t)\;
		|q_1\rangle_1 \otimes \cdots \otimes |q_{M}\rangle_{M},
	\end{equation}
	where $M \equiv N - n_0 + 1$ denotes the total number of modes, the summation runs over the associated Fock space of $M$ modes, and $|q\rangle_n$ denotes the photon-number state with $q$ photons in the mode indexed by $n$.  Under free-space evolution, the dynamics is governed by the Schr\"odinger equation $i\hbar\,\partial_t |\Psi(t)\rangle = \hat{H}|\Psi(t)\rangle$ with
	the free-field Hamiltonian
	\begin{equation}
		\hat H = \sum_{n=1}^{M}\hbar\,\omega_n\,\hat N_n,
		\label{eq:hamiltonian}
	\end{equation}
	where $\hat{N}_n=\hat{a}_n^\dagger \hat{a}_n$ is the photon-number operator associated with the
	$n$th mode, of frequency
	$\omega_n \equiv (2n_0 + 2n - 1)\omega$.  The time evolution of the general quantum state is therefore
	$|\Psi(t)\rangle
	= \sum_{\mathcal{F}}
	C_{q_1,\ldots,q_{M}}(0)\prod_{n=1}^{M}  |q_n\rangle_n\,
	\exp \bigl(-i\omega_n q_n t\bigr)$.
	
	\subsection{Field operators and observables}
	The central quantity of interest in our work is the electric-field-strength operator
	\begin{equation}\label{efield}
		\hat{\mathbf{E}}(\mathbf{r})
		= \sum_{n=1}^{M} \hat{\mathbf{E}}_n
		= i \sum_{n=1}^{M} \sum_{\mu}
		\mathbf{e}^{(\mu)}
		\sqrt{\frac{\hbar\omega_n}{2V\epsilon_0}}
		\Bigl(
		\hat{a}_n^{(\mu)}\,e^{i\mathbf{k}_n\cdot\mathbf{r}}
		- \hat{a}^{\dagger(\mu)}_n\,e^{-i\mathbf{k}_n\cdot\mathbf{r}}
		\Bigr).
	\end{equation}
	Since we consider identically polarised modes propagating along a single spatial direction, the polarisation vector $\mathbf{e}^{(\mu)}$ may be suppressed, and $\mathbf{k}_n \cdot \mathbf{r} = \omega_n x/c$ can be written. 
	We introduce the shorthand $\mathcal{E}_0 \equiv \sqrt{\hbar\omega / 2V\epsilon_0}$ interpreted as "electric-field strength per photon". 
	As is usual, we decompose the electric-field operator into its positive- and negative-frequency parts,
	$\hat{E} = \hat{E}^{(+)} + \hat{E}^{(-)}$, which, using the mode labelling from equation~\eqref{eq:hamiltonian}, are
	$\hat E^{(+)}(x) = i\mathcal{E}_0 \sum_{n=1}^{M}\sqrt{2n_0+2n-1}\,\hat a_n\,e^{i\omega_n x/c}$, and $\hat E^{(-)}(x) = \big(\hat E^{(+)}(x)\big)^{\dagger}$,
	satisfying the equal-time commutation relation
	$\big[\hat E^{(+)}(x_1),\hat E^{(-)}(x_2)\big] = 
	\mathcal{E}_0^2\sum_{n=1}^{M}(2n_0+2n-1) \exp\left(i\frac{\omega_n}{c}(x_1 - x_2) \right)$.

	Calculating the expectation value of the electric field is straightforward
	\begin{equation}\label{eq:mean-field}
		\langle\hat{E}(x,t)\rangle
		= i\mathcal{E}_0 \sum_{n=1}^{M} \sqrt{2(n_0+n)-1}
		\Bigl(
		\langle\hat{a}_n(t)\rangle\,
		\exp\left( i\omega_n x/c \right)
		- \langle\hat{a}^\dagger_n(t)\rangle\,
		\exp\left(-i\omega_n x/c \right)
		\Bigr),
	\end{equation}
	where $\omega_n = (2n_0+2n-1)\omega$ as defined in subsection (\ref{representationharmonic}). 
	Using the commutativity relation $[\hat E_n,\hat E_m]=0$ for $n\neq m$, the variance of the electric field admits the well-known decomposition
	\begin{equation}\label{eq:variance-decomp}
		\bigl(\Delta\hat{E}\bigr)^2
		= \Bigl(\Delta\sum_{n=1}^{M}\hat{E}_n\Bigr)^2
		= \sum_{n=1}^{M}\bigl(\Delta\hat{E}_n\bigr)^2
		+ 2\!\!\sum_{\substack{n,m=1\\n<m}}^{M}
		\operatorname{Cov}(\hat{E}_n,\hat{E}_m).
	\end{equation}
	The individual mode variances and (equal-time, equal-position) cross-mode covariances are given, respectively, by
	\begin{align}\label{eq:mode-variance}
		\bigl(\Delta\hat{E}_n (x,t)\bigr)^2
		&= -\mathcal{E}_0^2\,(2n_0+2n-1)
		\Bigl[
		\bigl(\langle\hat{a}_n^2(t)\rangle-\langle\hat{a}_n(t)\rangle^2\bigr)
		e^{2i\omega_n x/c}
		+\bigl(\langle\hat{a}_n^{\dagger2}(t)\rangle
		-\langle\hat{a}^\dagger_n(t)\rangle^2\bigr)
		e^{-2i\omega_n x/c}
		\nonumber\\
		&\quad
		-2\bigl(\langle\hat{a}^\dagger_n(t)\hat{a}_n(t)\rangle
		-\langle\hat{a}^\dagger_n(t)\rangle\langle\hat{a}_n(t)\rangle\bigr) - 1
		\Bigr],
	\end{align}\vspace{-0.3cm}
	\begin{align}\label{eq:mode-covariance}
		\operatorname{Cov}(\hat{E}_n(x,t),\hat{E}_m(x,t))
		= -\mathcal{E}_0^2\sqrt{(2n_0+2n-1)(2n_0+2m-1)} \hspace{5.3cm}
		\nonumber\\ 
		\quad\times\Bigl[
		\bigl(\langle\hat{a}_n(t)\hat{a}_m(t)\rangle
		-\langle\hat{a}_n(t)\rangle\langle\hat{a}_m(t)\rangle\bigr) e^{i(\omega_n+\omega_m)x/c}
		+\bigl(\langle\hat{a}^\dagger_n(t)\hat{a}^\dagger_m(t)\rangle
		-\langle\hat{a}^\dagger_n(t)\rangle\langle\hat{a}^\dagger_m(t)\rangle\bigr)
		e^{-i(\omega_n+\omega_m)x/c}
		\nonumber\\
		\qquad
		-\bigl(\langle\hat{a}_n(t)\hat{a}^\dagger_m(t)\rangle
		-\langle\hat{a}_n(t)\rangle\langle\hat{a}^\dagger_m(t)\rangle\bigr)
		e^{i(\omega_n-\omega_m)x/c}
		-\bigl(\langle\hat{a}^\dagger_n(t)\hat{a}_m(t)\rangle
		-\langle\hat{a}^\dagger_n(t)\rangle\langle\hat{a}_m(t)\rangle\bigr)
		e^{i(\omega_m-\omega_n)x/c}
		\Bigr].
	\end{align}

	Concerning the physical observation of quantised fields, it can be instructive to recall the process of photodetection by an ideal photon detector of negligible size (e.g., of atomic dimensions, with a frequency-independent photoabsorption cross-section) \cite{PhysRev.130.2529}. If such a detector is located at coordinate $x$, the probability per unit time that a photon is absorbed is proportional to $\langle \Psi(t)| \hat{E}^{(-)}(x) \hat{E}^{(+)}(x) | \Psi(t)\rangle$.
	More specifically, the expectation value associated with photo-detection by the ideal, small detector is
	\begin{equation}\label{eq:photodetection}
		\langle \hat{N}(x,t)\rangle \equiv
		G^{(1)}(x,t)=
		\sum_{n,m=1}^M \langle \hat{a}_n^\dagger(t) \hat{a}_m(t) \rangle 
		\exp\left(i\frac{\omega_m-\omega_n}{c}x\right).
	\end{equation}
	Local photodetection is sensitive to the intermodal correlations~\cite{RevModPhys.92.035005}.
	A photodetector much larger than the beam width, integrating over a time long compared with the pulse length, does not resolve the spatio-temporally fluctuating terms of equation~\eqref{eq:photodetection}.
	The appropriate operator associated with the total photodetection signal by a ``bucket detector'' reduces to the familiar total photon-number operator $\hat N = \sum_{n=1}^M\hat a_n^\dagger\hat a_n$.

	\subsection{Quantum optical description of pulse synthesis}\label{sec:synthesis}
	
	In classical models of pulse synthesis, mode-locked fields are optical beams whose spectrum consists of a frequency comb with well-defined phase relations between the complex amplitudes $\{\alpha_n\}$, associated with its constituent modes. Such mode-locked fields can contain very short periodic pulses of very high peak electric-field strength and, correspondingly, a large ratio of peak to time-averaged intensity, $E^2_{\rm peak}/E^2_{\rm average}$~\cite{FT92}. A closely related quantity is used in signal analysis~\cite{Nikookar_2013} to characterise peaks within a waveform: the peak-to-average power ratio $\eta$, equivalent to the square of the crest factor. This ratio provides a natural, quantitative measure of the temporal localisation of the electric field strength in a classical periodic pulse train. 
	However, there is no reason to restrict its use to the classical electric-field strength alone -- as we show below, its utility extends naturally to any Hermitian field operator, providing a basis for a genuinely quantum optical theory of pulse synthesis.
	
	\begin{figure}[h!]
		\centering
		\includegraphics[width=0.99\linewidth]{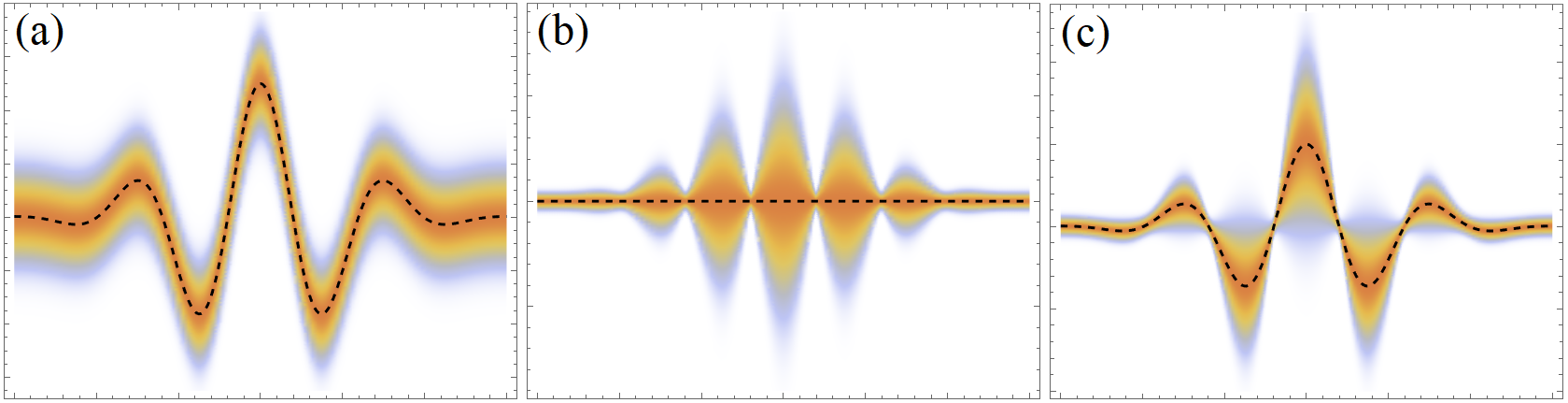}
		\caption{Illustration of the electric-field distribution within (a) a purely mean-field pulse, (b) a purely field-variance pulse, and (c) a general pulse, in which the field-variance maximum coincides with the field-average maximum. The classical value of the electric field is shown as a dashed line in each panel.}    
		\label{fig:attoquant2}
	\end{figure}
	
	The classical concept of ultrashort pulses can be naturally embedded into the domain of quantum optics, for example, in the following way: For an arbitrary Hermitian operator $\hat{O}$ constructed from the electric field of equation~\eqref{efield}, we define a peak-to-average power ratio
	\begin{equation}\label{etaOgeneral}
		\eta[\hat O] \;=\; \frac{\langle \hat O \rangle^{2}_{\mathrm{peak}}}{\langle \hat O \rangle^{2}_{\mathrm{average}}} ,
	\end{equation}
	where the maximum and average values are taken over a full spatiotemporal period.
	Equation~\eqref{etaOgeneral} quantifies the degree of localisation of the expectation value $\langle\hat O\rangle$, regardless of which statistical moment or cumulant of the field it represents. Cast in this form, the notion of an ultrashort pulse ceases to be tied exclusively to the classical field amplitude. A synthesis of an $\hat O$-related pulse train corresponds to achieving a high value of $\eta[\hat O]$. 

	It is instructive to consider the two types of pulse illustrated in figure~\ref{fig:attoquant2}(a) and (b). 
	We adopt the following nomenclature: temporal structures exhibiting a Fourier-locked electric-field expectation value arise from \emph{mean-field pulse synthesis}, whereas those exhibiting a Fourier-locked electric-field variance arise from \emph{field-variance pulse synthesis}.
	The former reduces, in the appropriate limit, to the classical description of mode-locked pulse trains, whilst the latter constitutes a genuinely quantum theoretical generalisation, with no direct classical counterpart.
	Both categories of pulses are captured within a single framework by $\eta[\hat E]$ and $\eta[(\Delta\hat E)]$, respectively. More generally, a hierarchy of pulse classes can be associated with the successive cumulants of the electric field: we refer to temporal structures with a Fourier-locked $N$th cumulant as resulting from \emph{field $N$th-cumulant pulse synthesis}. The corresponding pulses lie entirely outside of classical physics and become relevant whenever the state is significantly non-Gaussian.
	
	\subsubsection{Mean-field pulse synthesis by semiclassical states} \label{sec:field-average}
	
	Owing to its central importance in the classical description, we treat the class of separable products of coherent states separately. We introduce the notation
	$|\Phi_{\mathrm{SC}}(t)\rangle$ for these semiclassical quantum states:
	\begin{equation}\label{eq:semiclassical-state}
		|\Phi_{\mathrm{SC}}(t)\rangle
		= \prod_{n=1}^{M}
		\bigl| |\alpha_n|\,e^{-i\omega_n t + i\phi_n}\bigr\rangle_n\,
		e^{-i\omega_n t/2},
	\end{equation}
	where $\alpha_n = |\alpha_n|e^{i\phi_n}$ is the complex parameter associated with the coherent state in mode $n$, at $t=0$, with $|\alpha_n|\in\mathbb R$ being mode amplitudes and $\phi_n\in\mathbb R$ phases. We note that for a frequency-comb carrying a frequency-independent energy distribution across its constituent modes, 
	$\hbar \omega_n \langle \hat{a}^\dagger_n \hat{a}_n \rangle 
	= \hbar \omega_n |\alpha_n|^2$ must be independent of $n$, fixing $|\alpha_n|=C/\sqrt{2n_0+1+2n}$, with $C$ being an arbitrary real number. 
	Evaluating the electric-field expectation value reproduces the classical Fourier formula
	\begin{equation}\label{classicalEfield}
		\langle \hat{E}(x,t)\rangle_{\mathrm{SC}} =
		\langle\Phi_{\mathrm{SC}}(t)|\hat{E}(x)|\Phi_{\mathrm{SC}}(t)\rangle
		= -2\mathcal{E}_0 \sum_{n=1}^{M}
		\sqrt{2n_0+2n-1}\;|\alpha_n|
		\sin\!\Bigl(\omega_n\Bigl(\tfrac{x}{c}-t\Bigr)+\phi_n\Bigr).
	\end{equation}
	
	\begin{figure}[h!]
		\centering
		\includegraphics[width=0.995\linewidth]{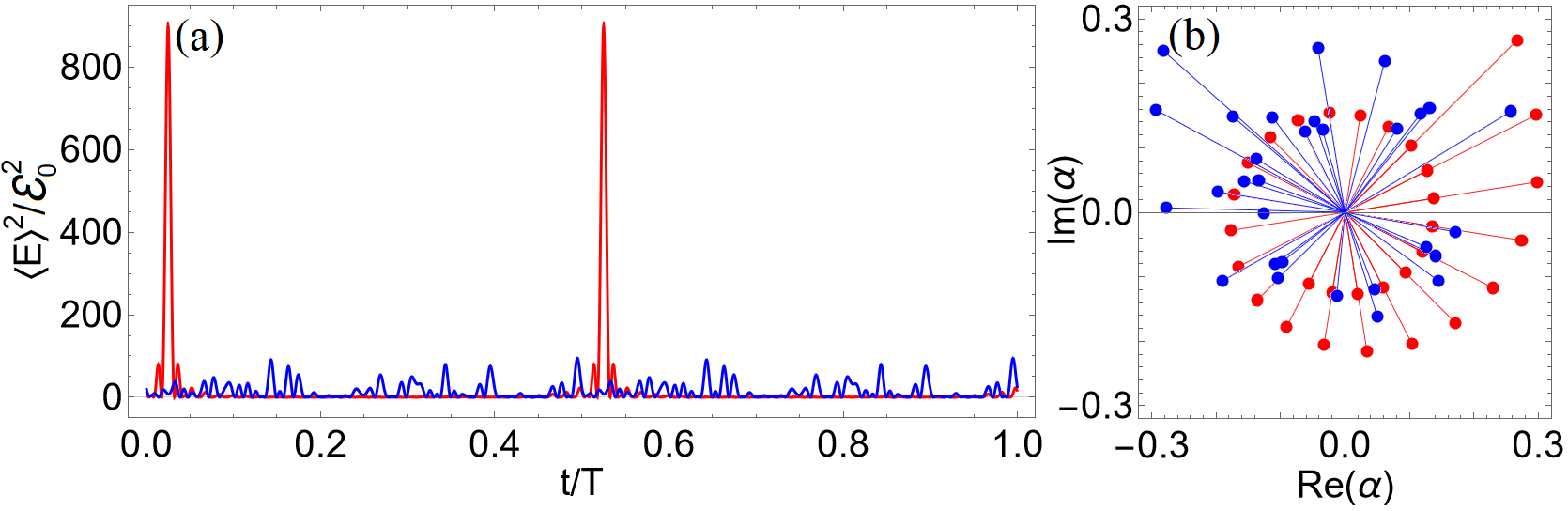}
		\caption{(a) Temporal structure of $\langle\hat E\rangle^2$ in units of $\mathcal{E}_0^2$, resulting from semiclassical states [equation~\eqref{eq:semiclassical-state}]. One can compare the cases of mode-locked (red) and random-phase (blue) conditions. (b) Coherent state parameters of the constituent coherent states in the complex plane, for the mode-locked (red) and random-phase (blue) cases.  Here M=30 harmonics (5th to 63rd) of identical harmonic energies are used, with $|\alpha_n|=\tfrac{1}{\sqrt{3+2n}}$.}
		\label{fig:attoquant1.png}
	\end{figure}
	The degree of temporal localisation, characterised by $\eta[\hat{E}]$, depends critically on the relative phases $\{\phi_n\}$.
	Maximal $\eta[\hat{E}]$ -- and consequently the most sharply localised pulses -- is reached in $(x,t)$ if the consecutive phase factors satisfy a uniform inter-mode phase difference
	$\phi_{n+1} - \phi_n
	= -2\omega\!\left(\frac{x}{c}-t\right)
	= \text{const}$. For calculational details, see Appendix \ref{appendixmodelocking}. This is precisely the mode-locking condition of classical attophysics.   
	Figure~\ref{fig:attoquant1.png} shows the contrast between the sharply peaked pulse train synthesised from mode-locked harmonics and the unstructured field resulting from a random phase distribution. 
	The order-of-magnitude difference between the resulting peaks of $\langle\hat E\rangle^2_{\rm SC}$ is evident.

	\subsubsection{Field-variance pulse synthesis by squeezed-vacuum product state}\label{sec:field-variance}
	
	A field-variance pulse train is a temporal structure within which the electric-field variance $(\Delta\hat E)^2$ exhibits periodic peaks, yielding a high peak-to-average ratio $\eta[(\Delta\hat E)]$. As an illustrative example, consider the squeezed-vacuum product state

	\begin{equation}\label{sqvacPsi}
		|\Phi_{\mathrm{SV}}(t)\rangle
		= \prod_{n=1}^{M}
		\bigl|r_n\,e^{i(\theta_n - 2\omega_n t)},\,0\bigr\rangle_n\,
		e^{-i\omega_n t/2},
	\end{equation}
	where $|z,0\rangle$ denotes the squeezed-vacuum state generated by the operator $\hat S(z)=\exp\big[(z^*\hat a^2-z\hat a^{\dagger 2})/2\big]|0\rangle$, with squeezing parameter $z \in \mathbb{C}$.
	Here, $\{r_n \in \mathbb{R}\}$ are the squeezing amplitudes and $\{\theta_n \in \mathbb{R}\}$ are the squeezing phases. 
	A frequency-independent energy distribution in the frequency-comb requires $\hbar \omega_n \langle \hat{a}^\dagger_n \hat{a}_n \rangle 
	= \hbar \omega_n \sinh^2(r_n)$ to be independent of $n$, implying $r_n=\text{arsinh}\left(C/\sqrt{2n_0+1+2n}\right)$, with $C$ being an arbitrary real number. 
	Since the intermodal covariances vanish, the electric-field variance reduces to a sum of single-mode contributions
	$  \bigl(\Delta\hat{E}\bigr)^2_{\mathrm{SV}}
	= \sum_{n=1}^{M}\bigl(\Delta\hat{E}_n\bigr)^2_{\mathrm{SV}}$,
	where the single-mode contributions are
	\begin{equation}
		\big(\Delta\hat E_n\big)^2_{\rm SV} = \mathcal{E}_0^2(2n_0+2n-1)\Big[1+2\sinh^2 r_n + \sinh(2r_n)\cos\Big(\frac{2\omega_n}{c}x-2\omega_n t+\theta_n\Big)\Big].
		\label{eq:squeezed-mode-variance}
	\end{equation}
	
	\begin{figure}[h!]
		\centering
		\includegraphics[width=0.992\linewidth]{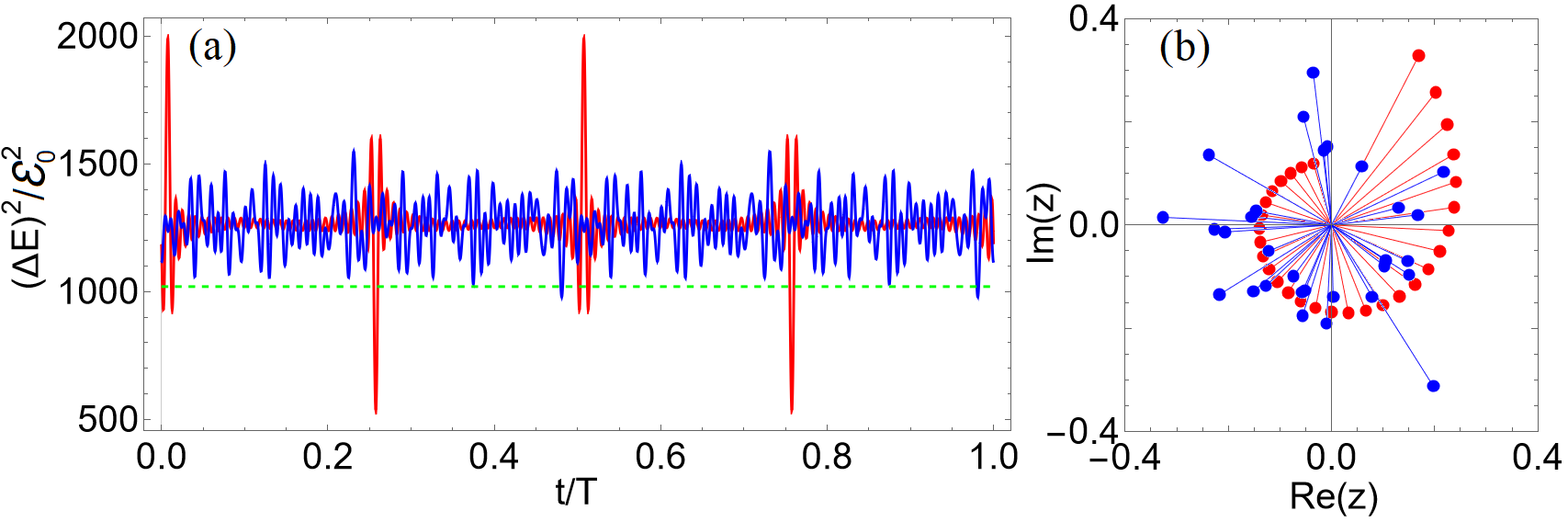}
		\caption{(a) Temporal structure of $(\Delta\hat{E})^2$ in units of $\mathcal{E}_0^2$, resulting from a product of squeezed-vacuum states [equation~\eqref{eq:squeezed-mode-variance}]. One can compare the cases of mode-locked (red) and random-phase (blue) conditions. The dashed green line indicates the corresponding vacuum variance. (b) Complex squeezing parameters of the constituent squeezed-vacuum states in the complex plane, for the mode-locked (red) and random-phase (blue) cases, respectively. Here M=30 harmonics (5th to 63rd) of identical harmonic energies are used, with $r_n=\text{arsinh}\tfrac{2.1}{\sqrt{2n_0+1+2n}}$.}
		\label{fig:attoquant3.png}
	\end{figure}
	The degree of temporal localisation $\eta[(\Delta \hat{E})_{\mathrm{SV}}]=(\Delta\hat{E})^2_{\mathrm{SV}\text{(peak)}}/(\Delta\hat{E})^2_{\mathrm{SV}\text{(average)}}$ again depends critically on the relative squeezing phases  $\{\theta_n\}$. Maximal $\eta[(\Delta \hat{E})_{\mathrm{SV}}]$ values are reached in $(x,t)$, if the difference of consecutive squeezing phase factors satisfies a relation of uniform difference
	$\theta_{n+1}-\theta_n = -4\omega(\tfrac{x}{c}-t) = \mathrm{const}$.
	For calculational details, see Appendix \ref{modelockingsqueezed}.
	This condition is the direct quantum counterpart to classical mode-locking for simple squeezed-vacuum product states. 
	Figure~\ref{fig:attoquant3.png} displays the contrast of the resulting field variance between pulse trains synthesised from mode-locked squeezing 
	and from a random distribution of squeezing phases. The difference in the resulting variance peaks, as well as in the presence of squeezing (in the sense of smaller variance than the corresponding multimode vacuum) is evident.

	\subsection{Necessity of mode-locking for special separable states}\label{sec:higher-moments} 
	As mentioned in subsection~(\ref{sec:synthesis}), the concept of pulse synthesis generalises in a straightforward manner to arbitrary electric-field-related quantities.
	The pulses resulting from the semiclassical states of~(\ref{sec:field-average}) and from the squeezed-vacuum product states of~(\ref{sec:field-variance}) represent the first two members of the field-cumulant pulse hierarchy. In both cases, mode-locking has been shown to be a necessary condition for optimising pulse-shapes.
	We now extend the discussion to arbitrary cumulants \cite{MW95} and establish the sense in which mode-locking is necessary for synthesising a train of maximally localised pulses from separable quantum states. To be clear, by necessity of mode-locking for pulse-shaping we mean that a train of maximally localised pulses requires a strict relationship between phases, i.e., a random choice of parameters will not yield any well-defined pulse structure.

	This necessity stems directly from the dependence of $\eta[\hat{O}]$ on the independent phase-parameters associated with distinct modes.
	In general, such independent phases need not exist at all; a simple counterexample for two modes is the two-mode squeezed state, whose single phase parameter characterises the mode pair jointly rather than either mode individually. Of course, the problem of defining a general quantum phase operator is involved even for a single mode, as is well known \cite{VS2008,Varro_2015}.
	To proceed, one needs well-defined phase-parameters. One-parameter unitary operators possess such phase-type quantities, namely the argument of their (nonzero) complex parameter.

	Consider a separable product state in which the $n$th mode is characterised by the one-mode state $|\psi_n\rangle_n = |\{\Lambda_{k,n}e^{i\varphi_{k,n}}\}, 0\rangle_n$. This one-mode state can be interpreted as the vacuum state acted upon by an otherwise unspecified product of unitary operators, characterised by the set of complex parameters $\{\Lambda_{k,n}e^{i\varphi_{k,n}}\}$ independent for each $(n,k)$ pair. Here $k$ indexes the distinct operators involved (e.g. displacement and squeezing operators) and $n$ is the mode index.
	One can interpret $\{\Lambda_{k,n}\}$ as a set of amplitudes associated with the $k$th operator affecting the $n$th mode, and $\{\varphi_{k,n}\}$ the corresponding set of phases. 
	Because the full state is a tensor product over modes, the phase parameters $\{\varphi_{k,n}\}$ of different modes are structurally independent of one another: nothing ties the phases of mode $n$ to those of any mode $n'\neq n$. 
	Consequently, the separable $M$-mode state defined above possesses at least $M$ independent phase-parameters, one attributable to each constituent mode, and more whenever a given one-mode component is generated by a product of distinct operators.
	It is this lower bound that forces the mode-locking condition. The two-mode squeezed state mentioned above illustrates the alternative: intermodal entanglement can push the number of independent phases below M, since here a single phase parameter characterises both modes jointly. 
	
	The spatiotemporal structure of the field cumulants, detailed in Appendix~\ref{cumulantappendix}, enters equation~\eqref{cumulantphase} through functions $f_{j,n,p}\left(\{\varphi_{k,n}\}\right)$. Here, mode-locking denotes a phase relationship --not necessarily uniform across modes-- that is required for pulse-shape optimisation, if these functions are genuinely independent for distinct values of $\omega' = p\omega_n$. Since a separable state carries at least one independent phase per mode, as established above, this independence holds generically throughout the family of separable states analysed in this section, and the classical necessity of mode-locking is recovered.
	Crucially, if $|\Psi\rangle$ is not a product state, the characteristic function no longer factorises as in equation~\eqref{eq:cumulant-additivity}; consequently, equation~\eqref{eq:cumulant-additivity-2} can fail and the ansatz of equation~\eqref{eq:Nth-order-ansatz} is no longer sufficient. By contrast with the separable case, the Fourier components of an entangled multimode state may share a smaller number of multimode phase parameters than there are modes; in extremal cases, all effective phases may become interdependent in such a way that the Fourier synthesis of the pulse becomes structurally locked. In the following section, we analyse a concrete quantum state that demonstrates this guaranteed locking of $\langle\hat{E}\rangle$ explicitly.
	
	

	\section{The symmetric entangled coherent state product}\label{sec:permanent}
	
	Explicit discussion of pulse synthesis in the previous section was limited to special separable states. 
	As noted, the classical correspondence between highly structured pulse trains and phase-locking can break down in entangled states. 
	In this section we introduce and characterise exactly such a state.
	
	\subsection{Definition}
	
	Consider a completely symmetrised entangled coherent state product, constructed as an equal superposition over all permutations of a fixed set of coherent-state amplitudes $\{\beta_k\}$ assigned to the $M$ harmonic modes. 
	This quantum state admits an elegant representation as a matrix permanent; hence, we will also call it a coherent permanent state:
	\begin{multline}\label{eq:perm-state-def}
		|\Psi(t)\rangle_{\mathrm{PERM}}
		=
		\frac{1}{\sqrt{\mathcal{N}}}
		\sum_{\sigma\in S_{M}}
		\prod_{n=1}^{M}
		\bigl|\beta_{\sigma(n)}\,e^{-i\omega_n t}\bigr\rangle_n
		\nonumber\\
		=
		\frac{1}{\sqrt{\mathcal{N}}}\,
		\mathrm{perm}
		\begin{pmatrix}
			\bigl|\beta_1\,e^{-i\omega_1 t}\bigr\rangle_1
			& \bigl|\beta_2\,e^{-i\omega_1 t}\bigr\rangle_1
			& \cdots
			& \bigl|\beta_M\,e^{-i\omega_1 t}\bigr\rangle_1
			\\[2pt]
			\bigl|\beta_1\,e^{-i\omega_2 t}\bigr\rangle_2
			& \bigl|\beta_2\,e^{-i\omega_2 t}\bigr\rangle_2
			& \cdots
			& \bigl|\beta_M\,e^{-i\omega_2 t}\bigr\rangle_2
			\\
			\vdots & & \ddots & \vdots
			\\
			\bigl|\beta_1\,e^{-i\omega_M t}\bigr\rangle_M
			& \bigl|\beta_2\,e^{-i\omega_M t}\bigr\rangle_M
			& \cdots
			& \bigl|\beta_M\,e^{-i\omega_M t}\bigr\rangle_M
		\end{pmatrix}.
	\end{multline}
	Here $\beta_k\in\mathbb{C}$ ($k=1,\ldots,M$) are the complex displacement parameters of the constituent coherent states, $S_M$ denotes the symmetric group on $M$ elements, and $\mathcal{N}$ is a normalisation constant determined below. 
	In this work, we suppress the trivial zero-point phase factors $e^{-i\omega_n t/2}$ associated with the vacuum energy of each mode.
	For completeness, we give the Fock-basis expansion coefficients $C_{q_1,\ldots,q_M}(t)$ defined in equation~(\ref{genquantumstate}) in Appendix \ref{appendixfockcoeff}. 
	This expansion suggests that when evaluating operator mean values, matrix permanents are generally expected to appear. Hence, the computational demand generally scales exponentially with the number of modes~\cite{scheel2004permanents}.
	One fundamental property of this state, at $t=0$, is its invariance under relabelling of modes by any permutation $\tau\in S_M$. Applying the mode-relabelling operator $\hat P_\tau$ (which maps $|\cdot\rangle_n\to|\cdot\rangle_{\tau(n)}$), one can check that
	\begin{equation}
		\hat P_\tau|\Psi(0)\rangle_{\rm PERM} = \frac{1}{\sqrt{\mathcal N}}\sum_{\sigma\in S_M}\prod_{n=1}^{M}\big(\beta_{\sigma(n)}\big)_{\tau(n)} = \frac{1}{\sqrt{\mathcal N}}\sum_{\sigma\in S_M}\prod_{n=1}^{M}\big(\beta_{(\sigma\circ\tau^{-1})(n)}\big)_n = |\Psi(0)\rangle_{\rm PERM},
		\label{eq:perm-invariance}
	\end{equation}
	where in the last step we used the fact that, as $\sigma$ ranges over $S_M$, so does $\sigma\circ\tau^{-1}$, leaving the sum unchanged. 
	
	Let us explicitly calculate the normalisation factor $\mathcal N$ here, to introduce the technique used throughout the paper. 
	\begin{equation}
		\mathcal N = \sum_{\sigma,\sigma'}\prod_{n=1}^{M} {}_n\!\big\langle \beta_{\sigma(n)}e^{-i\omega_n t}\big|\beta_{\sigma'(n)}e^{-i\omega_n t}\big\rangle_n = \sum_{\sigma,\sigma'}\prod_{n=1}^{M}\big\langle\beta_{\sigma(n)}\big|\beta_{\sigma'(n)}\big\rangle,
		\label{eq:norm-def}
	\end{equation}
	where the time-dependent phase factors cancel identically. We express the normalisation constant in terms of the Gram matrix of coherent-state overlaps. Define the $M\times M$ Hermitian Gram matrix $\mathbf{K}$ with elements
	\begin{equation}
		K_{pq} \equiv \langle\beta_p|\beta_q\rangle
		= \exp\!\Bigl[
		-\tfrac{1}{2}|\beta_p|^2
		-\tfrac{1}{2}|\beta_q|^2
		+\beta_p^*\beta_q
		\Bigr],
		\quad p,q = 1,\ldots,M.
	\end{equation}
	The normalisation constant is therefore
	\begin{equation}\label{Nraw}
		\mathcal{N}
		=
		\sum_{\sigma,\sigma'\in S_M}
		\prod_{n=1}^{M} K_{\sigma(n),\sigma'(n)}.
	\end{equation}
	To evaluate this double sum, fix $\sigma$ and re-index the product using $m=\sigma(n)$, and consequently $n=\sigma^{-1}(m)$, leading to
	$  \prod_{n=1}^{M} K_{\sigma(n),\sigma'(n)}
	=
	\prod_{m=1}^{M} K_{m,\,\sigma'(\sigma^{-1}(m))}
	=
	\prod_{m=1}^{M} K_{m,\,(\sigma'\circ\sigma^{-1})(m)}$.
	Defining $\tau = \sigma'\circ\sigma^{-1}$, note that as $\sigma'$
	ranges over $S_M$ with $\sigma$ fixed, $\tau$ identically ranges over $S_M$. Therefore
	$  \sum_{\sigma'\in S_M}
	\prod_{n=1}^M K_{\sigma(n),\sigma'(n)}
	=
	\sum_{\tau\in S_M}
	\prod_{m=1}^M K_{m,\tau(m)}
	=
	\mathrm{perm}(\mathbf{K})$,
	which is independent of $\sigma$. Summing over the $M!$
	permutations $\sigma$, we obtain the normalisation
	\begin{equation}\label{normalisation}
		\mathcal{N} = M!\;\mathrm{perm}(\mathbf{K}).
	\end{equation}
	With the quantum state explicitly expressed, we now turn toward its characterisation.

	\subsection{Photon statistics, Wigner function, and entanglement}
	To evaluate the photon statistics and to calculate the Wigner function associated with a single mode, we will make use of the reduced one-mode density matrix. The steps of tracing out all but the $n$th mode are detailed in Appendix \ref{densityM}. The resulting density matrix is
	\begin{equation}
		\rho_n = \frac{(M-1)!}{\mathcal N}\sum_{j,k=1}^{M}\mathrm{perm}\big(\mathbf\Omega_{k,j}\big)\,|\beta_j\rangle\langle\beta_k|,
		\label{eq:reduced-density-one-mode}
	\end{equation}
	where $\mathbf{\Omega}_{k,j}\in\mathbb C^{(M-1)\times(M-1)}$ is the matrix $\mathbf{K}$ with row $k$ and column $j$ deleted. Note that the reduced density matrix is independent of the mode index $n$, as anticipated from the permutation symmetry \eqref{eq:perm-invariance} of the state.

	\begin{figure}[h!]
		\centering
		\includegraphics[width=1.0\linewidth]{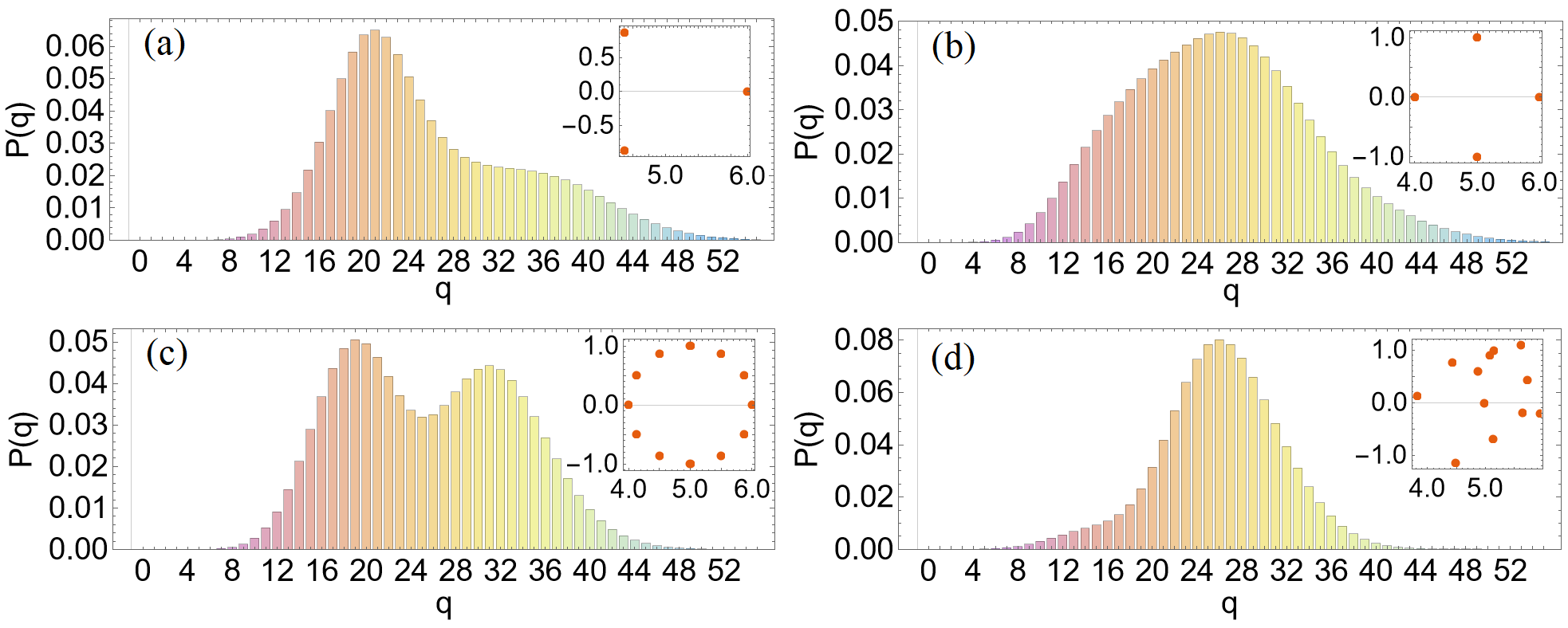}
		\caption{The photon statistics for $\{\beta_k=5+\delta_k\}$ parameters, shown on the inset of each subfigure. (a)-(c) correspond to $\delta_k$ evenly distributed on a unit circle, with M=3, 4 and 12 modes, respectively. (d) corresponds to randomly distributed $\delta_k$, with M=12 modes.}
		\label{fig:photonstats}
	\end{figure}
	
	The photon statistics of the $n$th mode can be calculated by $P(q_n)=\langle q | \rho_n | q\rangle$. Inserting equation~\eqref{eq:reduced-density-one-mode} into this formula, we obtain 
	\begin{equation}
		P(q_n)= \tfrac{(M-1)!}{\mathcal N}\!\sum_{j,k=1}^{M}\mathrm{perm} \big(\mathbf{\Omega}_{k,j}\big)\, \langle q |\beta_j\rangle\langle\beta_k| q \rangle
		=
		\frac{(M-1)!}{\mathcal N q!}\sum_{j,k=1}^{M}
		\mathrm{perm} \big(\mathbf{\Omega}_{k,j}\big)\, e^{-\tfrac{|\beta_j|^2+|\beta_k|^2}{2}} \left( \beta_k^* \beta_j\right)^{q}.
	\end{equation}
	The photon statistics, corresponding to a select combination of $\{\beta_k=5+\delta_k\}$ parameters, is shown on figure (\ref{fig:photonstats}). Here, $\delta_k$ is evenly distributed on a displaced unit circle in panels (a)–(c) of figure~\ref{fig:photonstats} and randomly distributed in panel (d). Deviations of the photon statistics from the Poissonian distribution are clearly visible. 
	
	
	\begin{figure}[h!]
		\centering
		\includegraphics[width=1.0\linewidth]{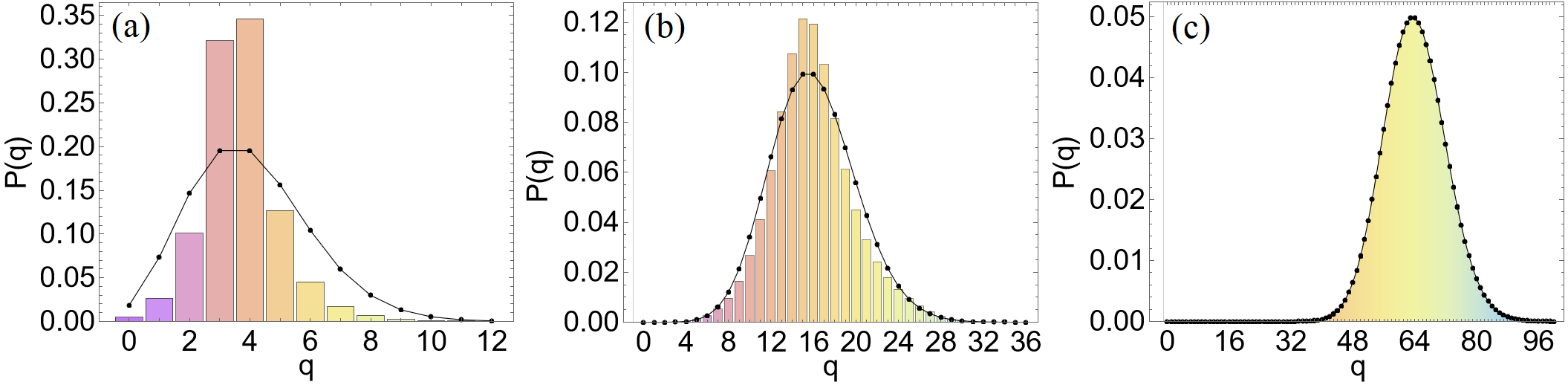}
		\caption{The photon statistics for $\{\beta_k\}$ parameters, distributed uniformly on a circle of modulus $|\beta_k|=2$ (a), $|\beta_k|=4$ (b), and $|\beta_k|=8$ (c). The corresponding Poissonian distribution is shown with black markers.}
		\label{fig:photonstats2}
	\end{figure}
	The photon statistics corresponding to a uniform angular distribution of $\{\beta_k\}$ with identical amplitude depends on this amplitude. On figure (\ref{fig:photonstats2}), one can observe that as the amplitude increases, the photon statistics transitions from a sub-Poissonian one to a Poissonian one. This can be interpreted as the interference between distinct coherent-state contributions disappearing as their parameters grow more distant.
	The Wigner function, in a similar manner, can be calculated based on equation~\eqref{eq:reduced-density-one-mode}, with the derivation given in Appendix \ref{wignerappendix}. The Wigner function of the coherent permanent states can be expressed as
	\begin{equation}\label{wignerfunction}
		W_{\rho_N}(\alpha) \;=\; \frac{2\,(M-1)!}{\pi\,\mathcal N}\sum_{j,k=1}^{M} K_{kj}\;\mathrm{perm}\bigl(\mathbf \Omega_{k,j}\bigr)\,\exp\!\left[-2(\alpha-\beta_j)(\alpha^*-\beta_k^*)\right] .
	\end{equation}
	
	\begin{figure}[h!]
		\centering
		\includegraphics[width=1.0\linewidth]{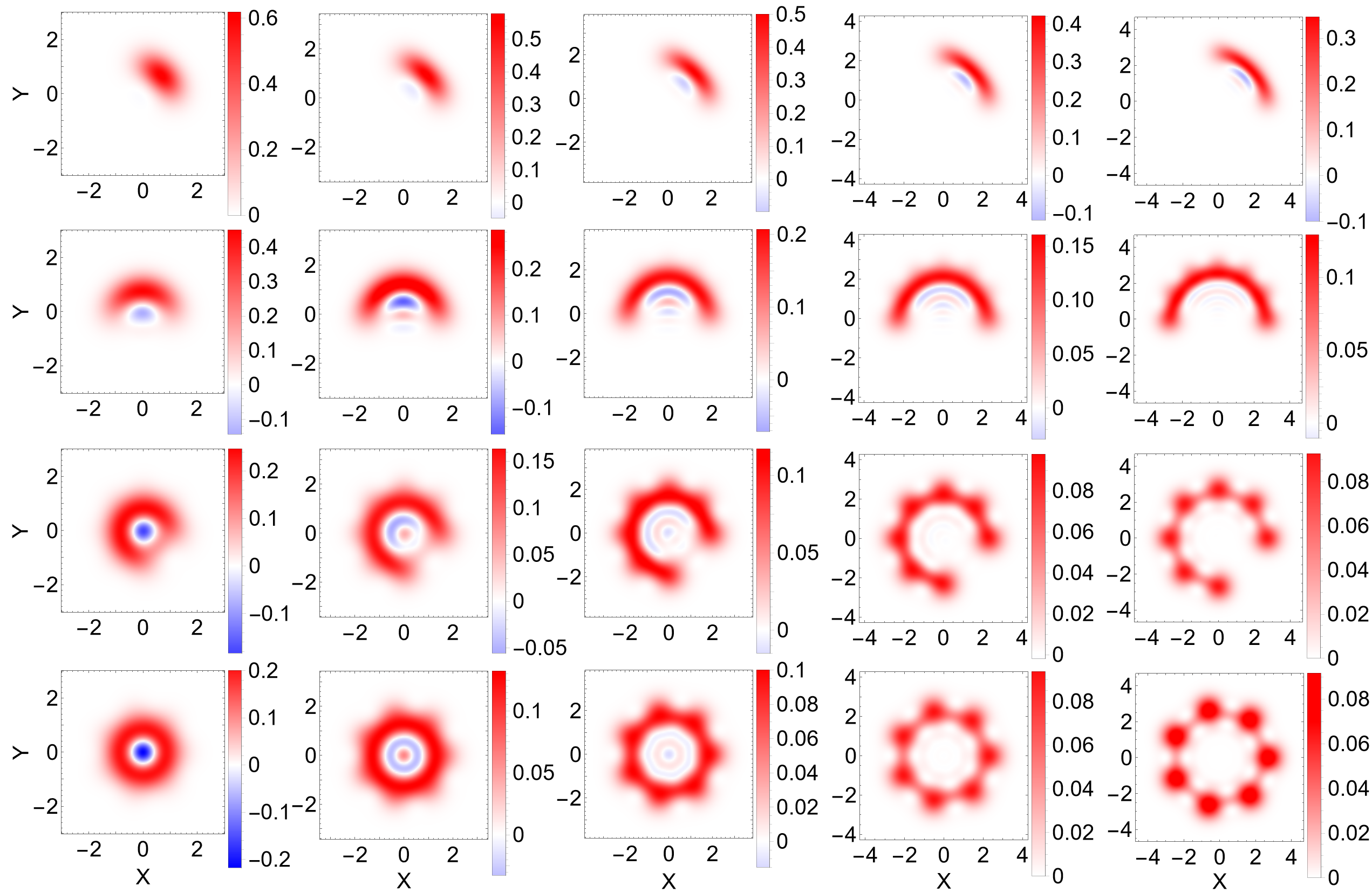}
		\caption{Wigner functions, calculated from equation~\eqref{wignerfunction} for $M=7$ modes, with $\{\beta_n\}$ evenly distributed along a quarter circle (first row), half-circle (second row), three-quarter circle (third row), and full circle (fourth row); of radii 1.1 (first column), 1.5 (second column), 1.9 (third column), 2.3 (fourth column) and 2.7 (fifth column).}
		\label{fig:wigner}
	\end{figure}
	The Wigner function is shown on figure (\ref{fig:wigner}), corresponding to $\{\beta_k\}$ parameters of identical modulus.
	One can observe the presence of interference within the Wigner function when the coherent state parameters are separated by an appropriate distance. Naturally, this interference is markedly different from that of coherent superpositions. 
	We highlight a subtlety for coherent-state parameters of identical modulus: although the Wigner function can display negative values, the resulting photon statistics may differ little from that of a coherent state with the corresponding amplitude.
	
	Going beyond the single-mode reduced density matrix, one can proceed with the characterisation of mode-pairs.
	Tracing out all but the $n$th and $m$th modes similarly gives the two-mode reduced density matrix, with details of the calculation given in Appendix \ref{densityNR}. The resulting density matrix is
	\begin{equation}
		\rho_{nm} = \frac{(M-2)!}{\mathcal N}\sum_{\substack{j,p=1\\j\neq p}}^{M}\sum_{\substack{k,q=1\\k\neq q}}^{M}\mathrm{perm}\big(\mathbf\Omega_{kq,jp}\big)\,\big(|\beta_j\rangle\langle\beta_k|\big)_n\otimes\big(|\beta_p\rangle\langle\beta_q|\big)_m,
		\label{eq:reduced-density-two-mode}
	\end{equation}
	where $\Omega_{kq,jp}\in\mathbb C^{(M-2)\times(M-2)}$ is the matrix $K$ with rows $k,q$ and columns $j,p$ deleted. Again, mode-permutation symmetry ensures that $\rho_{nm}$ is independent of the mode indices $n$ and $m$. 
	Thus, one can quantify the entanglement between any pair of harmonic modes via the logarithmic negativity of $\rho_{nm}$, derived in Appendix \ref{applogneg}.
	\begin{figure}[h!]
		\centering
		\includegraphics[width=1.0\linewidth]{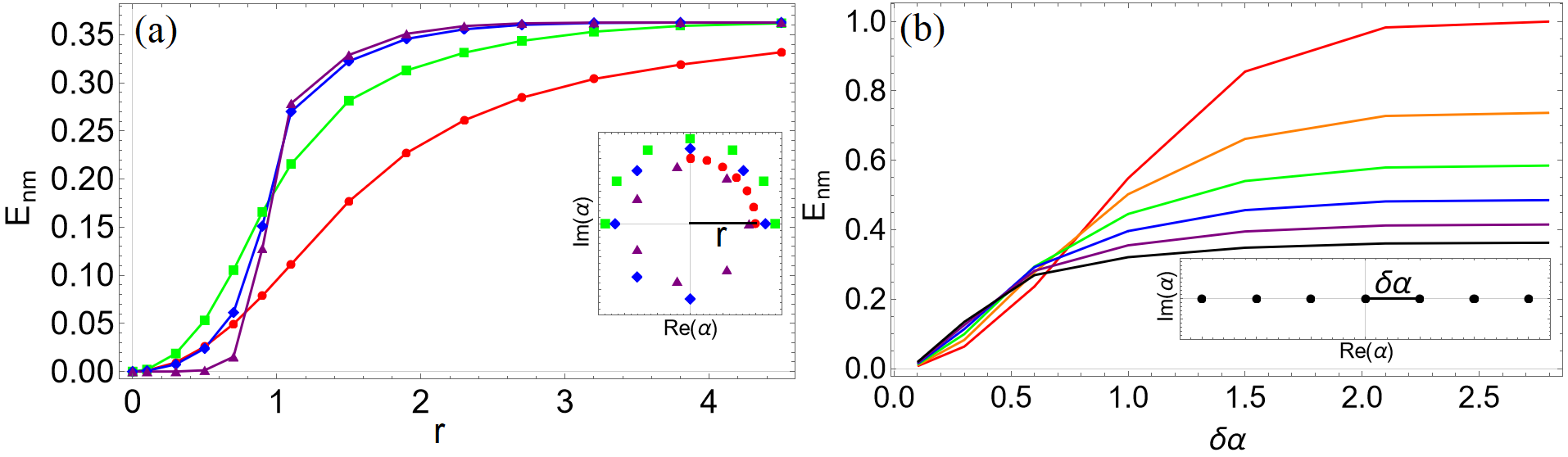}
		\caption{(a) Logarithmic negativity for $M=7$ modes, with the same type of distribution as in the rows of figure~(\ref{fig:wigner}),  as a function of the radius. Quarter-circle distribution is shown as red, half-circle as green, three-quarter circle as blue, full circle as purple. (b) Logarithmic negativity for symmetric linear distribution, with $\delta\alpha$ distance between the nearest parameters. The mode number is M=2 shown as red, M=3 as orange, M=4 as green, M=5 as blue, M=6 as purple, M=7 as black.}
		\label{fig:logneg}
	\end{figure}
	As one can see on figure~(\ref{fig:logneg}), the logarithmic negativity increases as the distances between the coherent-state parameters increase, until it reaches an asymptotic value. By the point at which the interferences in the Wigner function disappear, the asymptotic value of the logarithmic negativity is closely approximated. The asymptotic value depends primarily on the total mode number $M$, and decreases with it, as shown on figure~\ref{fig:logneg}(b). At the same time, the rate at which the logarithmic negativity increases depends on the exact distribution of $\{\beta_n\}$. One can observe on figure~\ref{fig:logneg}(a) that when this distribution is uniform on a circle of given modulus, the logarithmic negativity can jump suddenly to near its asymptotic value as $r$ increases. Note that the near-Poissonian photon statistics and the near-asymptotic logarithmic negativity can coexist. This has implications concerning potential experimental identification of such quantum states, which we will not dwell on here. 
	
	While these calculations characterise the state's internal statistical properties, they do not directly reveal the behaviour of the electromagnetic field. To relate this state to the pulse synthesis discussed earlier, we must first evaluate the fundamental moments of the field. 
	
	\subsection{One and two-mode normal-ordered expectation values}
	We calculate the general one-mode normal-ordered expectation value $\langle\hat{a}^{\dagger k}_n \hat{a}^{l}_n\rangle$. 
	Using the same calculational technique as used for the normalisation, we obtain
	\begin{equation}\label{adagkal}
		\Bigl\langle \hat{a}^{\dagger k}_n \hat{a}^l_n \Bigr\rangle
		=
		\frac{e^{i(k-l)\omega_n t}}{M\cdot\mathrm{perm}(\mathbf{K})}
		\sum_{j=1}^{M} \beta^{*k}_j \mathrm{perm}\bigl(\mathbf{K}^{[j]}_{l}\bigr),
	\end{equation}
	where $\mathbf{K}^{[j]}_{l}$ denotes the $\mathbf{K}$ matrix with row $j$ replaced element-wise by $(\beta_1^l K_{j,1}, \dots \beta_M^l K_{j,M})$. 
	For details of the calculation, see Appendix \ref{sec:one-mode}.
	Naturally, $\mathbf{K}^{[j]}_{0}=\mathbf{K}$, hence, it is easy to check that in the special cases of $l=0$ or $k=0$, the results can be greatly simplified. Specifically,
	\begin{equation}\label{adagkexp}
		\langle\hat{a}^{\dagger k}_n\rangle
		=
		e^{i k \omega_n t}
		\sum_{j=1}^{M} \frac{\beta^{*k}_j}{M} 
		=
		\mathbb{E}(\beta^{*k}) e^{i k \omega_n t} ,
		\hspace{0.4cm} \text{and similarly} \hspace{0.4cm}
		\langle\hat{a}^{l}_n\rangle
		=
		\mathbb{E}(\beta^{l}) e^{-i l \omega_n t} .
	\end{equation}
	Here, we used the complex-valued average $\mathbb{E}(\beta^k) \equiv \frac{1}{M}\sum_{j=1}^{M}\beta_j^k$.
	Observe that the mode index $n$ appears in equation~(\ref{adagkal}) only in the phase-factor $e^{i(k-l)\omega_n t}$. 
	The photon-number expectation value, on the other hand, does not yield a closed-form expression in general, and has no simple interpretation in terms of the $\{\beta_n\}$ parameters:
	\begin{equation}\label{eq:photonmean}
		\langle \hat{N}_n \rangle=
		\langle \hat{a}^{\dagger}_n \hat{a}_n \rangle
		=
		\frac{1}{M\cdot\mathrm{perm}(\mathbf{K})}
		\sum_{j=1}^{M} \beta^{*}_j \mathrm{perm}\bigl(\mathbf{K}^{[j]}_{1}\bigr) .
	\end{equation}
	While the one-mode moments are sufficient to determine the mean-field behaviour, other properties of the field, such as the field variance, are influenced by the inter-harmonic correlations. 
	
	We calculate the general two-mode expectation value
	$\langle\hat{a}^{\dagger k}_n\hat{a}^l_n\hat{a}^{\dagger f}_m\hat{a}^g_m\rangle$
	for distinct $n\neq m$ modes. Following the steps of Appendix \ref{twomodeexpectation}, the result reads
	\begin{equation}\label{twomode}
		\Bigl\langle\hat{a}^{\dagger k}_n\hat{a}^l_n
		\hat{a}^{\dagger f}_m\hat{a}^g_m\Bigr\rangle
		=
		\frac{e^{i[(k-l)\omega_n+(f-g)\omega_m]t}}
		{M(M-1)\,\mathrm{perm}(\mathbf{K})}
		\sum_{\substack{j,j'=1\\j\neq j'}}^{M}
		\beta^{*k}_j\,\beta^{*f}_{j'}\,
		\mathrm{perm}\bigl(\mathbf{K}^{[j,j']}_{l,g}\bigr),
	\end{equation}
	where $\mathbf{K}^{[j,j']}_{l,g}$ is $\mathbf{K}$ with row $j$ replaced by $\bigl(\beta^l_v K_{j,v}\bigr)_v$ and row $j'$ replaced by $\bigl(\beta^g_v K_{j',v}\bigr)_v$.
	Similarly to the one-mode quantities, the results simplify considerably in the special cases $l=g=0$ or $k=f=0$.
	For $k=f=0$, $\mathbf{K}^{[j,j']}_{l,g}$ carries no $\beta^*$ weight on rows $j$ and $j'$, so that
	\begin{equation}
		\big\langle\hat a_n^l\hat a_m^g\big\rangle = \frac{e^{-i(l\omega_n+g\omega_m)t}}{M(M-1)}\sum_{\substack{j,j'=1\\j\neq j'}}^{M}\beta_j^{l}\,\beta_{j'}^{g}.
		\label{eq:two-mode-annihilation-building-block}
	\end{equation}
	Since 
	$\sum_{\substack{j,j'=1;j\neq j'}}^{M}\beta_{\tau(j)}^{l}\,\beta_{\tau(j')}^{g} = \Big(\sum_j \beta_{\tau(j)}^l\Big)\Big(\sum_{j'}\beta_{\tau(j')}^g\Big) - \sum_j\beta_{\tau(j)}^{l+g} = M^2 \mathbb{E}(\beta^l)\mathbb{E}(\beta^g) - M\,\mathbb{E}(\beta^{l+g})$
	holds for any bijection $\tau$, we can obtain the closed-form expressions
	\begin{equation}
		\big\langle\hat a_n^l\hat a_m^g\big\rangle = \frac{e^{-i(l\omega_n+g\omega_m)t}}{M-1}\Big[M\,\mathbb{E}(\beta^l)\mathbb{E}(\beta^g) - \mathbb{E}(\beta^{l+g})\Big],
		\label{eq:two-mode-annihilation2}
	\end{equation}
	\begin{equation}
		\big\langle\hat a_n^{\dagger k}\hat a_m^{\dagger f}\big\rangle = \frac{e^{i(k\omega_n+f\omega_m)t}}{M-1}\Big[M\,\mathbb{E}(\beta^{*k})\mathbb{E}(\beta^{*f}) - \mathbb{E}(\beta^{*(k+f)})\Big].
		\label{eq:two-mode-creation2}
	\end{equation}
	Here, the second line follows by Hermitian conjugation of the first.
	Further special cases of importance are the terms $\langle\hat a_n^\dagger\hat a_m\rangle$ for $n\neq m$, that is,\ $k=1, l=0, f=0, g=1$ in equation~\eqref{twomode}. These terms do not yield a simple closed-form expression analogous to equations~\eqref{eq:two-mode-annihilation2}--\eqref{eq:two-mode-creation2} but can be rewritten in a useful way. After some calculation (for details, see Appendix~\ref{mixedcorr}), it reads
	\begin{equation}\label{eq:mixed-covariance-closed}
		\langle\hat a_n^\dagger\hat a_m\rangle = \frac{e^{i(\omega_n-\omega_m)t}}{M-1} \left[  M|\mathbb{E}(\beta)|^2-\langle\hat N_n\rangle \right] \qquad n\neq m.
	\end{equation}
	Thus, calculating the first-order intermodal correlator is no more involved than calculating the mean photon number itself: beyond $\langle\hat N_n\rangle$, no further permanent evaluation is required.
	
	With these results, every one- and two-mode moment entering the electric-field variance and covariance, equations~\eqref{eq:mode-variance}--\eqref{eq:mode-covariance}, is expressed through $\mathbb E(\beta)$ and the single permanent-valued quantity $\langle\hat N_n\rangle$. Higher moments such as $\langle\hat a_n^{\dagger 2}\hat a_n^2\rangle$ or $\langle\hat N_n\hat N_m\rangle$ generally require the  evaluation of a permanent.

	\subsection{Spectral energy distribution}
	We close this characterisation with a small note on the mean photon numbers.
	The value of $\langle\hat N_n\rangle$ in equation~\eqref{eq:photonmean} carries no explicit mode dependence. Hence, for a coherent permanent state on $M$ modes, the mean photon numbers are identical for each mode. Consequently, the energy distribution among modes, $\hbar\omega_n\langle\hat N_n\rangle\propto(2n_0+2n-1)$ grows linearly with the harmonic index $n$, irrespective of the ($\mathbf K$-dependent) value of $\langle\hat N_n\rangle$ itself. This differs from typical frequency combs, where the spectral intensity is roughly constant whilst the photon number per mode decreases with frequency: here it is the photon number per mode that forms a plateau, whilst the intensity grows linearly with the harmonic order.


	\section{The coherent permanent states as attoquants}\label{attoquant}
	
	To be clear and unambiguous: We define the term ``attoquants'' or ``attoquant states'' as the set of multimode quantum states of the harmonic field that synthesise a robust $\langle\hat{E}\rangle$-pulse train. By robustness of the pulse train, we mean that $\eta\big[\hat{E}\big]$ is independent of the individual phase-parameters.
	The singular term ``attoquant'' denotes an individual multimode quantum state satisfying this condition.
	
	One route to construct an attoquant, the one taken in section~\ref{sec:permanent}, is to take a symmetrised superposition over all permutations of individual modes. Such a construct renders the quantum state automatically invariant to the exchange of any two modes (ignoring the mode-dependent time evolution). It also leads to the state being a matrix permanent, with the matrix elements corresponding to the constituent single-mode states. 
	Within this general construct, the single-mode constituents need not be coherent states at all: an attoquant can, in principle, contain different types of single-mode excitations.
	We conjecture, but do not prove here, that more general attoquants may be built from e.g. displaced number states.
	Here, we limit ourselves to showing that the completely symmetric entangled coherent-state product of section~\ref{sec:permanent} is indeed an attoquant.

	\subsection{Electric field expectation value}\label{meanfieldattoquant}
	
\begin{figure}[h!]
		\centering
		\includegraphics[width=0.95\linewidth]{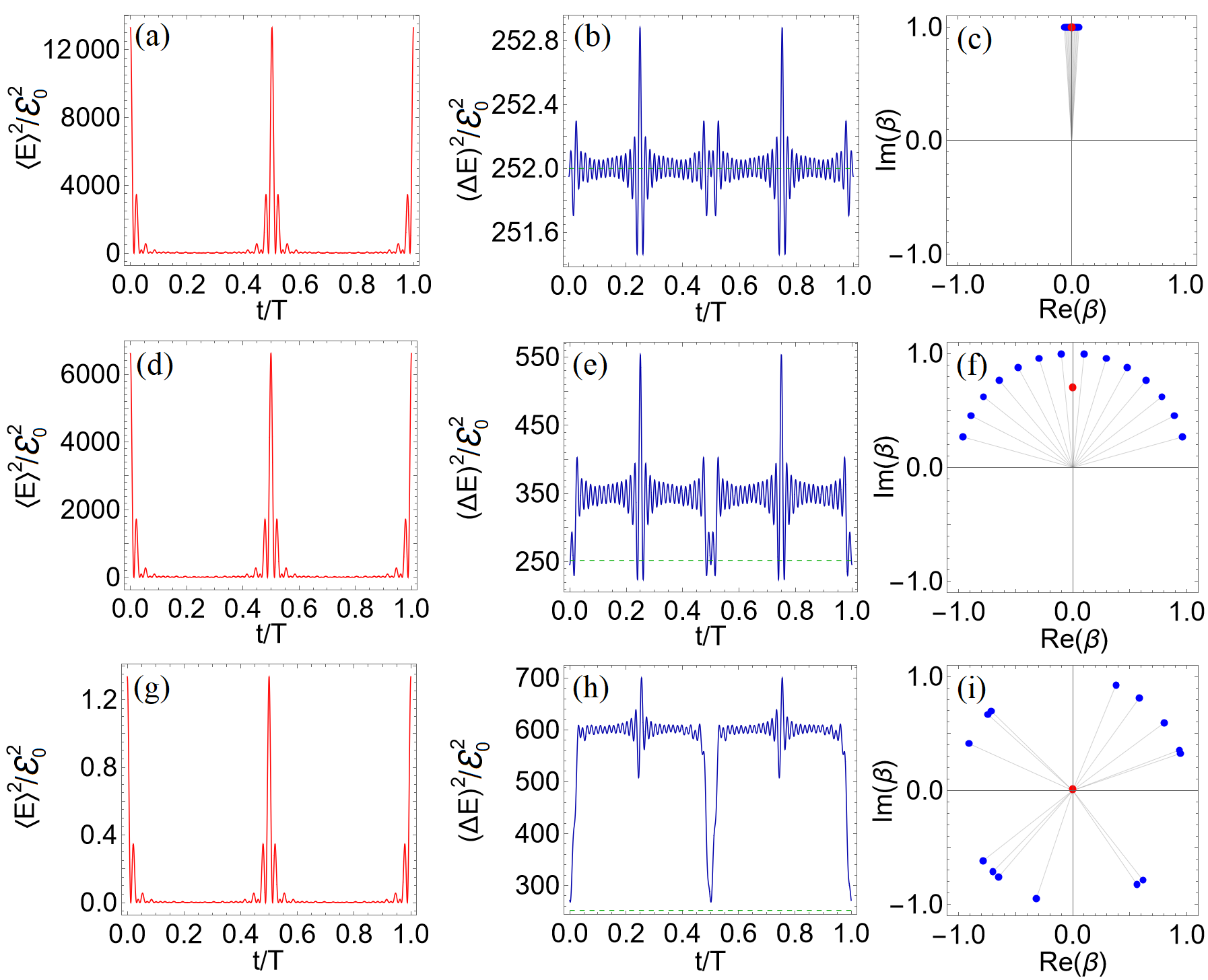}
		\caption{Spatiotemporal structure of the field average (a), (d), (g) and variance (b), (e), (h), for mode-locked and closely aligned parameters (c) in the first row, mode-locked and less aligned parameters (f) in the second row, and random parameters (i) in the third row. All parameters are of unit modulus, and their averages are shown with red markers on (c), (f), (i).}
		\label{fig:meanandvariance}
	\end{figure}

	Let us start by noting that the expectation value of any weighted sum of creation and annihilation operators is straightforward to calculate from the results of the previous section. This is a natural direction of approach, since the mean electric field-strength $\langle\hat{E}\rangle$ is itself just a particular choice of weights. It follows directly from equation~\eqref{adagkexp} that 
	\begin{equation}\label{sum}
		\langle u_1 \hat{a}_1 + u_2 \hat{a}_2 + \dots + u_M \hat{a}_M\rangle_{\mathrm{PERM}}=
		(u_1 e^{-i\omega_1 t} +u_2 e^{-i\omega_2 t}+\dots+u_M e^{-i\omega_M t})
		\frac{\beta_1 + \beta_2 +\dots + \beta_M }{M}.
	\end{equation}
	The dependence of this expression on the individual $\{\beta_k\}$ is carried exclusively by the single $\mathbb{E}(\beta)$ coefficient. The resulting Fourier locking is therefore independent of the complex parameters $\{\beta_k\}$ for any choice of weights. This locking is universal in any propagation direction, which is one of the most important results of this work. The resulting pulse-train structure can be illustrated by considering the $u_1=u_2=\dots=u_M=u$ special case, for which the temporal structure on the right-hand side of equation~\eqref{sum} is proportional to a Dirichlet-kernel-like expression $\sum_{n=1}^{M}e^{i\omega_n t}=e^{i\omega t(2n_0+M)}\tfrac{\sin\left(M\omega t \right)}{\sin(\omega t)}$, characteristic of frequency combs.
	Substituting equation~\eqref{adagkexp} into the field expectation value of equation~\eqref{eq:mean-field} gives the mean-field value
	\begin{equation}\label{eq:field-mean-perm}
		\langle\hat{E}(\tau)\rangle_{\mathrm{PERM}}
		=
		-2\mathcal{E}_0\,|\mathbb{E}(\beta)|
		\sum_{n=1}^{M}\sqrt{2n_0+2n-1}\;
		\sin \bigl(\omega_n\tau+\varphi_0\bigr), \qquad \tau\equiv\frac xc-t,
	\end{equation}
	where we introduced the spatiotemporal $\tau$ parameter for simplicity. Here, $\varphi_0 = \arg\mathbb{E}(\beta)$ is the phase of the average $\{\beta_k\}$ parameters. Since every Fourier component shares this common effective phase, the resulting superposition is locked. 
	Provided $\mathbb E(\beta)\neq0$, the superposition synthesises a pulse for \emph{any} distribution of the individual phases, including for random distributions, in sharp contrast to the semiclassical state of subsection~\ref{sec:field-average}, see figure~\ref{fig:meanandvariance}(a), (d) and (g).
	One can readily check that $\eta[\hat{E}]$ takes a fixed value, insensitive to the phases of $\{\beta_k\}$.
	This robustness is precisely the defining feature of attoquants. We have therefore shown that the coherent permanent state is indeed an attoquant.
	This is, moreover, a genuinely novel quantum optical feature, stemming directly from the symmetry properties of the quantum state $|\Psi(t)\rangle_{\mathrm{PERM}}$. 

	\subsection{Variance of the electric field}
	
	First, let us define two shorthand quantities that let us express equations~\eqref{eq:variance-decomp}--\eqref{eq:mode-covariance} more succinctly: the complex variance of the displacement parameters, $\mathrm{Var}_{\mathbb C}(\beta)\equiv\mathbb E(\beta^2)-\mathbb E(\beta)^2~(\in\mathbb C)$, and the photon-number deviation, $\langle\delta\hat{N}\rangle\equiv\langle\hat N_n\rangle-|\mathbb E(\beta)|^2$.
	The single-mode variance of equation~\eqref{eq:mode-variance} can be expressed using equation~\eqref{adagkexp}, resulting in $\langle\hat a_n^2\rangle-\langle\hat a_n\rangle^2=\mathrm{Var}_{\mathbb C}(\beta)e^{-2i\omega_nt}$. Together with $\langle\hat a_n^\dagger\hat a_n\rangle-\langle\hat a_n^\dagger\rangle\langle\hat a_n\rangle=\langle\delta\hat{N}\rangle$, it reads
	\begin{equation}\label{eq:single-mode-variance-full}
		\bigl(\Delta\hat E_n(\tau)\bigr)^2_{\rm PERM} = \mathcal E_0^2(2n_0+2n-1)\Bigl[1+2\langle\delta\hat{N}\rangle-2\,\mathrm{Re}\bigl(\mathrm{Var}_{\mathbb C}(\beta)\,e^{2i\omega_n\tau}\bigr)\Bigr].
	\end{equation}
	The variance oscillates with $2\omega_n$ about the mean value $\mathcal E_0^2(2n_0+2n-1)(1+2\langle\delta\hat{N}\rangle)$, the oscillation amplitude determined by $\mathrm{Var}_{\mathbb C}(\beta)$.
	The static excess over the vacuum value $\mathcal E_0^2(2n_0+2n-1)$ is governed by $\langle\delta\hat{N}\rangle\geq0$ alone (its non-negativity is proved in Appendix~\ref{appDeltaN}). 
	The oscillation itself vanishes if and only if $\mathrm{Var}_{\mathbb C}(\beta)=0$. This holds whenever $\beta_1=\cdots=\beta_M$, but not only then; $M$ points spaced evenly around a circle can also satisfy the $\mathrm{Var}_{\mathbb C}(\beta)=0$ condition.
	
	The four two-mode terms entering equation~\eqref{eq:mode-covariance} follow from equations~\eqref{eq:two-mode-annihilation2}-\eqref{eq:mixed-covariance-closed} and are
	\begin{equation}\label{eq:cov-summary}
		\begin{aligned}
			\langle\hat a_n\hat a_m\rangle-\langle\hat a_n\rangle\langle\hat a_m\rangle &= -\frac{\mathrm{Var}_{\mathbb C}(\beta)}{M-1}\,e^{-i(\omega_n+\omega_m)t}, &
			\langle\hat a_n^\dagger\hat a_m^\dagger\rangle-\langle\hat a_n^\dagger\rangle\langle\hat a_m^\dagger\rangle &= -\frac{\mathrm{Var}_{\mathbb C}(\beta)^*}{M-1}\,e^{i(\omega_n+\omega_m)t}, \\
			\langle\hat a_n^\dagger\hat a_m\rangle-\langle\hat a_n^\dagger\rangle\langle\hat a_m\rangle &= -\frac{\langle\delta\hat{N}\rangle}{M-1}\,e^{i(\omega_n-\omega_m)t}, &
			\langle\hat a_n\hat a_m^\dagger\rangle-\langle\hat a_n\rangle\langle\hat a_m^\dagger\rangle &= -\frac{\langle\delta\hat{N}\rangle}{M-1}\,e^{-i(\omega_n-\omega_m)t}.
		\end{aligned}
	\end{equation}
	Substituting these terms into equation~\eqref{eq:mode-covariance} gives the intermodal field covariances as
	\begin{equation}\label{eq:field-cov-closed}
		\mathrm{Cov}(\hat E_n,\hat E_m) = \frac{2\mathcal E_0^2\sqrt{(2n_0+2n-1)(2n_0+2m-1)}}{M-1}\Bigl\{\mathrm{Re}\bigl[\mathrm{Var}_{\mathbb C}(\beta)\,e^{i(\omega_n+\omega_m)\tau}\bigr]-\langle\delta\hat{N}\rangle\cos\bigl[(\omega_n-\omega_m)\tau\bigr]\Bigr\}.
	\end{equation}
	Collecting the single-mode contributions $\sum_n(\Delta\hat E_n)^2$ from equation~\eqref{eq:single-mode-variance-full} 
	and the covariances above gives the expression for the total variance of the electric field in a coherent permanent state (for details see Appendix~\ref{sumvari}) as
	\begin{align}\label{eq:total-variance-perm}
		\bigl(\Delta\hat E(\tau)\bigr)^2_{\rm PERM} = \mathcal E_0^2 M(M+2n_0) 
		+ \frac{2\mathcal E_0^2}{M-1}
		\left\{\langle\delta\hat{N}\rangle
		\left[M^2(M+2n_0)- \left| \sum_{n=1}^M\sqrt{2n_0+2n-1}\,e^{i\omega_n\tau} \right|^2
		\right] \right.
		\nonumber\\
		\left.
		+ \mathrm{Re}\left[\mathrm{Var}_{\mathbb C}(\beta)
		\left(\left( \sum_{n=1}^M\sqrt{2n_0+2n-1}\,e^{i\omega_n\tau} \right)^2 -M\,\sum_{n=1}^M (2n_0+2n-1)\,e^{2i\omega_n\tau} \right)\right]\right\}.
	\end{align}
	The variance separates into a flat vacuum floor $\mathcal E_0^2M(M+2n_0)$ and two quantum contributions. The first, proportional to $\langle\delta\hat{N}\rangle$, is non-negative (proved in Appendix~\ref{appDeltaN}). The second, oscillating and proportional to $\mathrm{Var}_{\mathbb C}(\beta)$, can be negative, and can therefore push the variance locally below the vacuum floor.
	
	Because the variance depends on the full second-moment structure of $\{\beta_k\}$ rather than on $\mathbb E(\beta)$ alone, its spatiotemporal structure is not protected against dephasing, as illustrated in figure~\ref{fig:meanandvariance}(b), (e), (h). However, since the mode energies $\hbar\omega_n\langle\hat N_n\rangle$ are fixed by $\langle\hat N_n\rangle$ rather than by $\mathbb E(\beta)$, configurations with a suppressed mean field ($\mathbb E(\beta)\approx0$, e.g.\ from random phases) typically show enhanced $\langle\delta\hat{N}\rangle$ instead. In this qualitative sense, suppressing the mean-field pulse through dephasing can reinforce the field-variance pulse instead.
	
	
	
	\subsection{Photodetection by an ideal detector}

	Let us briefly discuss the highly ideal, small-detector photodetection signal of equation~\eqref{eq:photodetection},
	\begin{equation} 
		G^{(1)}(x,t) = \sum_{n,m=1}^M\langle\hat a_n^\dagger(t)\hat a_m(t)\rangle\,e^{i(\omega_m-\omega_n)x/c}.
	\end{equation}
	The sum can be split into a time-independent diagonal, and oscillating off-diagonal terms. Evaluation with respect to an attoquant is trivial from equation~\eqref{eq:photonmean} and from equation~\ref{eq:mixed-covariance-closed}, and reads
	\begin{equation}\label{eq:photonumber-xt} 
		G^{(1)}(x,t) = 
		M\langle N_n \rangle 
		+
		\dfrac{M \left| \mathbb E(\beta) \right|^2 - \langle N_n \rangle}{M-1}
		\sum_{n\neq m} e^{i(\omega_m -\omega_n)\tau}. 
	\end{equation}
	The first term of equation~\eqref{eq:photonumber-xt} is the time-averaged signal, measured by photodetectors with large bandwidth integrating over a time-interval larger than the driving field's optical cycle. The second term shows that an ideal detector could, in principle, register the pulse-structure of the attoquant state.
	
	\subsection{Special case: Coherent W-state-like configuration}\label{sec:special}
	
	The results of sections~\ref{sec:permanent} and~\ref{attoquant} show that closed-form, transparent expressions for various quantities are generally out of reach, owing to the permanent structure that inevitably appears. Here, we examine a special case that is particularly transparent for analysis.
	Let $\beta_1=\alpha$ differ from the remaining $M-1$ parameters $\beta_k=\beta_0$ ($k=2,\ldots,M$), which are all identical. This configuration can be considered a symmetrised deformation away from the semiclassical state of section~\ref{sec:field-average}. The Gram matrix is then unity almost everywhere, with the first row and column being the sole exceptions. Let us introduce the notation $\kappa\equiv \langle\alpha|\beta_0\rangle=\exp\Big[-\tfrac12|\alpha|^2-\tfrac12|\beta_0|^2+\alpha^*\beta_0\Big]$, with which the Gram matrix is
	\begin{equation}
		\mathbf{K}
		=
		\begin{pmatrix}
			1 & \kappa & \cdots & \kappa \\
			\kappa^* & 1 & \cdots & 1 \\
			\vdots        & \vdots        & \ddots & \vdots        \\
			\kappa^* & 1 & \cdots & 1
		\end{pmatrix}.
	\end{equation}
	A straightforward calculation gives $
	\mathrm{perm}(\mathbf{K}) = (M-1)!\,\bigl[1+(M-1)|\kappa|^2\bigr]$ and hence $
	\mathcal{N} = M!\,(M-1)!\,\bigl[1+(M-1)|\kappa|^2\bigr]$,
	where $|\kappa|^2=e^{-|\alpha-\beta_0|^2}$. In addition, one can express $\mathbb E(\beta^l) = \frac{\alpha^l+(M-1)\beta_0^l}{M}$ and $\mathrm{Var}_{\mathbb C}(\beta) = \frac{M-1}{M^2}(\alpha-\beta_0)^2$.
	Consequently, the mean-field amplitude vanishes when $\alpha=-(M-1)\beta_0$.
	The variance of the electric field, together with other calculations, is given in Appendix~\ref{specialcase}. 
	
	\begin{figure}[h!]
		\centering
		\includegraphics[width=0.9\linewidth]{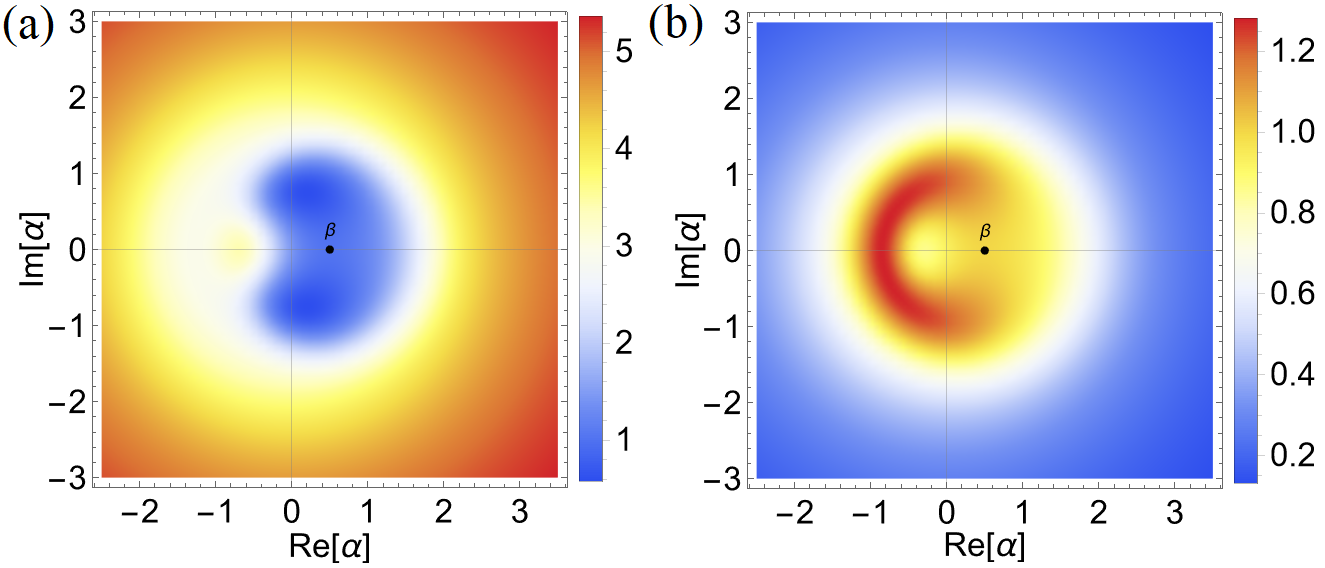}
		\caption{Normalised second-order correlation functions, with $\beta_0=1/2$ and $M=7$ fixed, as a function of $\alpha$. (a) shows $g_n^{(2)}$ of equation~\eqref{eq:special-A-g2-one}, while (b) shows $g_{nm}^{(2)}$ of equation~\eqref{eq:special-A-g2-two}.}
		\label{fig:corrfunc}
	\end{figure}
	This special state of parameters also permits direct access to the state's nonclassical photon statistics, which we characterise through its second-order correlation functions.
	Dividing equation~(\ref{eq:special-A-adag2a2}) by $\langle\hat N_n\rangle^2$, using equation~\eqref{specialcasemeanN} for $\langle\hat N_n\rangle$, yields the one-mode second-order correlation function in closed form:
	\begin{equation}\label{eq:special-A-g2-one}
		g^{(2)}_n
		=
		\frac{M\bigl[1+(M-1)|\kappa|^2\bigr]\,
			\Bigl\{|\alpha|^4+2(M-1)|\kappa|^2\,\mathrm{Re}[(\alpha^*\beta_0)^2]
			+(M-1)|\beta_0|^4[1+(M-2)|\kappa|^2]\Bigr\}}
		{\Bigl\{|\alpha|^2+2(M-1)|\kappa|^2\,\mathrm{Re}(\alpha^*\beta_0)
			+(M-1)|\beta_0|^2[1+(M-2)|\kappa|^2]\Bigr\}^{2}}.
	\end{equation}
	In the well-separated limit $|\kappa|^2\to0$ with $|\beta_0|\ll |\alpha|$, $g^{(2)}_n\to M$, and the quantum state becomes strongly super-bunched.
	The two-mode second-order correlation can be calculated by dividing equation~\eqref{eq:special-A-NnNm} by $\langle\hat N_n\rangle\langle\hat N_m\rangle=\langle\hat N_n\rangle^2$,
	\begin{equation}\label{eq:special-A-g2-two}\hspace{-0.8cm}
		g^{(2)}_{nm}
		=
		\frac{M\bigl[1+(M-1)|\kappa|^2\bigr]\,|\beta_0|^2
			\Bigl\{2|\alpha|^2(1+|\kappa|^2)+4(M-2)\,\mathrm{Re}(\alpha^*\beta_0)\,|\kappa|^2
			+(M-2)|\beta_0|^2[1+(M-3)|\kappa|^2]\Bigr\}}
		{\Bigl\{|\alpha|^2+2(M-1)|\kappa|^2\,\mathrm{Re}(\alpha^*\beta_0)
			+(M-1)|\beta_0|^2[1+(M-2)|\kappa|^2]\Bigr\}^{2}}.
	\end{equation}
	Numerical evaluation across the parameter space shows that $g^{(2)}_n$ takes higher values, corresponding to super-Poissonian statistics, for most choices of $\alpha$. Sub-Poissonian statistics appear only for a small region of parameter space, as seen in figure~\ref{fig:corrfunc}(a). This is in sharp contrast with the circularly distributed parameters, the photon statistics of which are shown in figure~\ref{fig:photonstats2}. The two-mode correlation $g^{(2)}_{nm}$ can have values either above or below unity, depending on $M$ and the relative values of $\alpha$ and $\beta_0$. Values below unity are typical, with $g^{(2)}_{nm}>1$ confined to a smaller region of parameter space, as seen in figure~\ref{fig:corrfunc}(b). High one-mode and high intermodal correlations can coincide in some region of parameters.

	\section{Discussion}\label{sec:dicsussion}
	
	Having established the mathematical properties of the coherent permanent state, we now turn to a discussion of its physical relevance. First, we outline an open question concerning the quantum optics of HHG. Second, we propose a physical picture for how attoquants might arise. Finally, we discuss how the concept of attoquants can help answer the questions raised.

	\subsection{An open question}
	
	The broader question we start with is the following: why should the collective response of an assembly of essentially independently radiating atoms reproduce the same attosecond temporal structure usually associated with a single isolated emitter? 
	The phase-locking condition, applied to the dipole-moment of a single atom~\cite{antoine1997phase,PhysRevLett.77.1234} --i.e. constant relative phase-difference between subsequent dipole-components-- by itself, says nothing about the spatial coherence or directionality of the radiation, observed experimentally. Those properties are instead governed by the macroscopic phase-matching conditions for propagation through the extended gas medium, as developed extensively by L'Huillier and co-workers \cite{PSalieres_1996}. 
	A first-principles description connecting the microscopic origin of the dipole phase with the macroscopic propagation picture remains an open problem.
	Our aim here is 
	to offer a quantum-optical framework in which the temporal locking of the high-harmonic components emerges naturally.

	Before we clarify what constitutes the fundamental puzzle more specifically, let us recall some important aspects of HHG. 
	The ``standard semiclassical procedure'' of decomposing the induced atomic dipole moment into a Fourier series,
	\begin{equation}
		\mathbf{d}(t) = -e\langle\psi(t)|\hat{\mathbf{x}}|\psi(t)\rangle = \sum_n \mathbf{d}_n\, e^{-in\omega_0 t},
		\label{eq:dipole}
	\end{equation}
	is equivalent, term by term, to a description in which every harmonic mode of the quantised radiation field is driven by spontaneous emission from its vacuum state to a one-photon state~\cite{ehlotzky1992harmonic,varro1993generation,varro1993new,photonics8070269}. 
	The equivalence between the two pictures can be made explicit at the level of first-order perturbation theory. 
	For a single atom, the relevant matrix element of the field operator, taken between the vacuum and the one-photon state of a mode $\omega'$, is simply
	$\langle\Phi_f|\hat{\mathbf{E}}(\mathbf{r},t')|\Phi_i\rangle \;\sim\; \langle 1_{\omega'}|\hat{A}^+_{\omega'}|0_{\omega'}\rangle = 1$,
	so that the amplitude for emitting one quantum into that mode is
	\begin{equation}
		T_{fi} = \frac{ie}{\hbar}\int_0^t dt'\, \langle\psi_f|\hat{\mathbf{x}}(t')|\psi_i\rangle\cdot\langle\Phi_f|\hat{\mathbf{E}}(\mathbf{r},t')|\Phi_i\rangle,
		\label{eq:Tfi}
	\end{equation}
	driven by the same dipole matrix element that appears in the classical Larmor-type radiation term in equation~\eqref{eq:dipole}. This connection was already noted in early quantum treatments of high-harmonic generation, including the 1990 work of Becker et al.~\cite{PhysRevA.41.4112}, who considered a delta-function model potential.
	Although the process as a whole is driven by the strong fundamental field, the emission into each initially-empty harmonic mode is itself spontaneous, and it is far from obvious why harmonics generated this way should lock in phase. While the driven dipole's phase is perfectly well defined, fixed by the local driving field, the timing of the individual spontaneous-emission events is genuinely stochastic, following an exponential decay law. 
	Therefore, the uncertainty of the instant at which a given harmonic photon is emitted introduces stochastic quantum noise into the harmonic phases. Naively summing many such independent harmonic emission events would ordinarily be expected to degrade any fixed phase relation between harmonics, rather than reinforce it. It follows that from a quantum-optical perspective, it is a genuine puzzle that harmonics can combine into a phase-locked frequency comb.

	\subsubsection{Dependence of the quantum state on atomic position}
	
	Of course, equation~\eqref{eq:Tfi} is only the weak-signal limit of a fully non-perturbative, quantised treatment. A combination of squeezing and displacement transformations can eliminate the minimal-coupling interaction terms, and give a much better estimate of the harmonic amplitudes. Calculations within the quantised Kramers-Henneberger formalism~\cite{varro1993new,photonics8070269} show that the amplitudes are governed primarily by the matrix elements of a multiphoton effective potential
	\begin{equation}
		V_{\hat{\boldsymbol\alpha}} = V(\mathbf{r}+\hat{\boldsymbol\alpha}) - V(\mathbf{r}),\qquad
		V(\mathbf{r}+\hat{\boldsymbol\alpha}) = \int d^3q\,\widetilde{V}(\mathbf{q})\,\exp\!\Big[\frac{i}{\hbar}\mathbf{q}\cdot(\mathbf{r}+\hat{\boldsymbol\alpha})\Big],
		\label{eq:Va}
	\end{equation}
	where the displacement by operator $\hat{\boldsymbol\alpha}\propto (\hat a^{+}-\hat a)$ stems from the dynamical displacement of the electron’s position.  Since this operator appears in the exponent on the right-hand side of equation~\eqref{eq:Va}, it generates a displacement acting on the fundamental mode. Extending the formulae in \cite{photonics8070269} to the complete set of harmonic modes, we naturally arrive at the product of displacement operators generating the coherent state product. Moreover, as already noted in \cite{photonics8070269}, all mode operators are multiplied by phase factors of the type $e^{-i\textbf{kR}}$ where $\textbf{R}$ is the position of the nucleus. Extending these formulae to the many-mode, many-atom case, we have operator$  ~\hat{\boldsymbol\alpha}\propto a_n^\dagger e^{-i\textbf{k}_n\textbf{R}_j} - a_n e^{i \textbf{k}_n\textbf{R}_j}$,
	where $\textbf{R}_j$ denotes the position of the $j$th gas atom in the jet. This results in a product of quantum optical displacement operators containing the geometric arrangement of the emitters.
	
	Between dressed states $|\psi_{0}\rangle|n_0\rangle$ and $|\psi_{1}\rangle|n_1\rangle$ of the atom and the fundamental mode, the corresponding second-order transition amplitude takes the form~\cite{varro1993new,photonics8070269}
	\begin{equation}
		T_{1,0}\ \propto\ \delta\{\hbar\omega' - [(n_0-n_1)\hbar\omega + E_0-E_1]\}\,\big(A_{1,0}+B_{1,0}\big),
		\label{eq:KH}
	\end{equation}
	\begin{equation}
		A_{1,0} = \sum_r \frac{\langle\psi_1|(\hat{\mathbf p}\cdot\boldsymbol\varepsilon')|\psi_r\rangle\,\langle\psi_r|\langle n_1|V_{\hat{\boldsymbol\alpha}}|n_0\rangle|\psi_0\rangle}{E_r-E_0-(n_0-n_1)\hbar\omega+i0},
		\label{eq:A10}
	\end{equation}
	\begin{equation}
		B_{1,0} = \sum_r \frac{\langle\psi_1|\langle n_1|V_{\hat{\boldsymbol\alpha}}|n_0\rangle|\psi_r\rangle\,\langle\psi_r|(\hat{\mathbf p}\cdot\boldsymbol\varepsilon')|\psi_0\rangle}{E_r-E_0+\hbar\omega'+i0},
		\label{eq:B10}
	\end{equation}
	where $|\psi_0\rangle,|\psi_1\rangle$ are the atomic states before and after the process, $E_0,E_1$ their energies, and $\boldsymbol\varepsilon'$ the polarisation vector. The sum over $r$ runs over the complete set of intermediate (bound and continuum) states. The two terms $A_{1,0}$ and $B_{1,0}$ correspond to the two possible time-orderings of the process: absorption of $(n_0-n_1)$ photons from the fundamental mode, followed by emission of the harmonic photon, or the reverse.
	While these formulae demonstrate how the atomic positions influence the quantised process, they cannot, by themselves, address what happens once the elementary processes they describe are repeated, essentially independently, by every atom in the illuminated volume. It is to this many-atom problem, and its implications with respect to symmetric entangled states, that we now turn.

	\subsection{A possible origin of permanent coherent states}
	
	High-harmonic generation in a gas-jet target is essentially a nonlinear light-scattering process. It is therefore useful to briefly revisit a key aspect of the quantum theory of light scattering by atoms.
	Let us focus first on the simplest case: the scattering of a single quantum from two spatially separated atoms. The coherence of the scattered field has a transparent physical origin, identified long ago by Heitler in the context of resonance fluorescence \cite{Heitler1944}. Let the distance between the two atoms be denoted by $\mathbf{R}$. Classically, the waves scattered from the two atoms simply interfere, and the resulting interference pattern is fixed entirely by their relative phase, determined by the path-length difference set by $\mathbf{R}$ and the scattering directions. Heitler's argument was that this classical picture remains valid in the fully quantised theory, for a specifically quantum-theoretical reason: a measurement of the scattered light, even at the level of individual photon counts, is generally unable to reveal which of the two atoms did the scattering. Since this information is absent from the quantum state of the field, the two histories cannot be assigned separate probabilities. Instead of considering them as mutually exclusive events, these scattering histories must be superposed coherently. 
	
\begin{figure}[h!]
		\centering
		\includegraphics[width=1.0\linewidth]{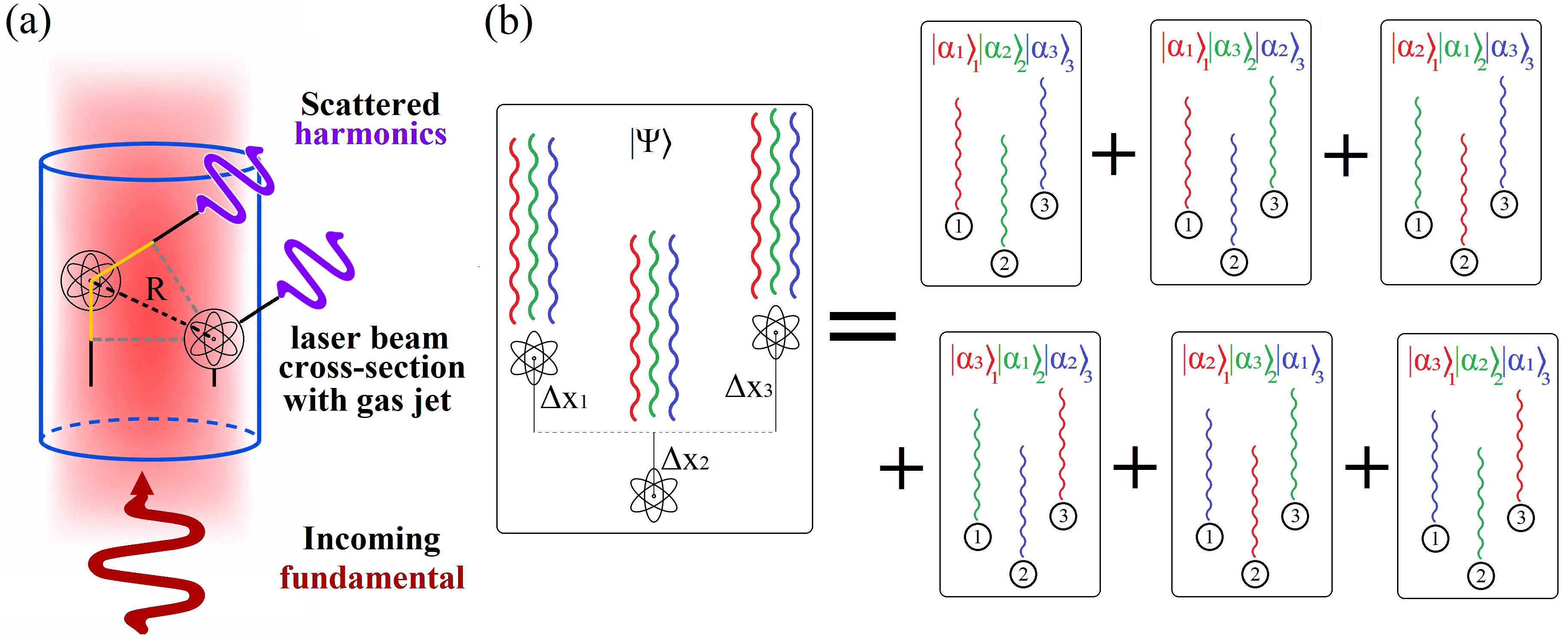}
		\caption{(a) Physical picture underlying the coherent permanent state: a focused fundamental beam (red) illuminates a coherence cluster of atoms across the gas-jet cross-section; each atom re-radiates a comb of high harmonics (purple), whose path length difference (yellow) to a common far-field detector depends on the atomic positions. (b) For a minimal cluster of $3$ atoms at longitudinal positions $\Delta x_1,\Delta x_2,\Delta x_3$, each is nominally associated with one of three coherent-state amplitudes ($\alpha_1,\alpha_2,\alpha_3$) for the three harmonic modes (red, green, blue). Since nothing distinguishes which atom actually fed which mode, the true state $|\Psi\rangle$ is the symmetric superposition of all $3!=6$ ways of assigning these three amplitudes to the three modes: the attoquant.}
		\label{fig:photonemission}
	\end{figure}

	An important ingredient renders this two-atom argument relevant to understanding the many-atom effects influencing HHG: the separation of length scales. The driving (fundamental) field occupies a coherence volume $\sim\lambda_0^3$, essentially corresponding to the interaction volume, whereas the coherence volumes of its harmonics, $\lambda_q^3=\lambda_0^3/q^3$ for the $q$th harmonic, are much smaller. While the driving excites a coherence cluster of atoms collectively, the harmonics it induces are radiated individually. On the much finer scale of the high-order harmonic wavelength, the atoms become spatially resolvable. This is precisely the regime in which Heitler's path-length argument is non-trivial, since $\mathbf{R}$ must be compared with the scattered, not the driving, wavelength. 
	This suggests a specific microscopic picture for a coherence cluster of $M$ atoms, illustrated for $M=3$ in figure~\ref{fig:photonemission}. In a given elementary event of the kind described by equations~\eqref{eq:KH}-\eqref{eq:B10}, one atom may emit photons in any of the $M$ harmonic modes, but each atom emits only one harmonic photon at a time. Conversely, a given harmonic photon may originate from any of the $M$ atoms in the cluster. Because no observable of the emitted light records which spatially resolved atom is responsible for which harmonic photon, the different possible assignments have to be superposed coherently at the level of the field state, exactly as in Heitler's two-atom argument. Of course, now for a full permutation of $M$ objects rather than a single transposition of two. Let us be clear about what is, and is not, being symmetrised here. The atoms themselves are treated as classically distinguishable objects, individually localised -- well within their own de Broglie wavelength -- at the resolved longitudinal positions $\Delta x_j$ shown in figure~\ref{fig:photonemission}(b). Hence, we do not symmetrise an atomic many-body wavefunction under particle exchange. What must be symmetrised is the assignment of field amplitudes to modes.

	Concretely, suppose that atom $j$ ($j=1,\dots,M$), following a single-atom dynamics, would populate the harmonic modes with a product of coherent states. The complex $\{\alpha_j\}_{n=1}^M$ parameter of a given harmonic mode is influenced by the driving field's local amplitude and phase at the location of the $j$th atom. Even more importantly, the $\{\alpha_j\}_{n=1}^M$ parameters carry the propagation phases $e^{i k_n \Delta x_j}$ of the emitted harmonic. These are obviously influenced by the atom's position $\Delta x_j$, and correspond to the same path-length differences as those present in Heitler's argument. The state of the $M$-mode field emitted by the whole cluster is then the equally-weighted superposition over all atom-to-mode assignments $\sigma$ in the symmetric group $S_M$,
	\begin{equation}
		|\Psi\rangle = \tfrac{1}{\mathcal{N}} \sum_{\sigma\in S_M} \bigotimes_{n=1}^{M} \big|\alpha_{\sigma(n)}\big\rangle_n .
		\label{eq:permanent}
	\end{equation}
	This sum over every permutation of $M$ mode indices assigned to $M$ atoms is the same combinatorial structure as the coherent permanent state analysed in the previous sections. This is the reason why we call the many-mode quantum states of this combinatorial character attoquants. For $M=3$, this state is shown expanded into its $3!=6$ terms in figure~\ref{fig:photonemission}(b). A coherence cluster containing more than $M$ atoms can be accounted for by restriction to the $M$ atoms that happen to feed a chosen set of modes. We do not develop this generalisation further here. 

	The equal-weight superposition of equation~\eqref{eq:permanent} deserves some justification. In an elementary emission process, one atom generates one harmonic photon: although atom $j$ can, in principle, be displaced into any of the $M$ harmonic modes, and thus has the capacity to emit any of the harmonics, simultaneous emission of several harmonic photons by the same atom is a negligible-probability process. The driving field couples identically to every atom, while this is not necessarily so for the harmonic  modes. The atoms are indistinguishable with respect to which one feeds a given harmonic mode's occupation. This indistinguishability does not, by itself, imply that the individual atomic contributions should be summed coherently into a single amplitude $\sum_j\beta_j$ as could be the case during e.g. superradiance: coherent addition of the displacement parameters would be expected only if all atoms lay within a single coherence volume of the highest-order harmonic involved. At typical gas-jet densities, however, the atoms are classically well separated and, at the scale of high-harmonic coherence length, do not form a single, directly coupled quantum system. Each atom has an equal claim to contribute to a given harmonic mode in a position-dependent way, and it is this equality claim -- realised as a superposition over the classical permutations of the atoms -- that is captured by the permanent structure. 

An immediate question can be raised with respect to the domain of validity of the strict permutation symmetry. Indeed, it is strictly valid only when the interaction volume within the gas jet is small compared to the wavelength of the driving field. On the other hand, inhomogeneity of the field amplitude across the beam does not, by itself, alter the permanent structure of equation~\eqref{eq:permanent}. It would, of course, affect the variance in $\beta$ values, which are dependent on the local field experienced by each atom.
Dispersion during the propagation in a medium could produce a distinct effect, deteriorating the phase-locking. We do not expect this to be significant in gas-jet targets, due to propagation length being small, and the frequency dependence of the refractive index being weak in gases at the frequencies of interest. 

\subsubsection{Directionality of attoquant radiation}

Recall that the electric-field-strength operator of equation~\eqref{efield} was considered with an arbitrary one-dimensional propagation in equation~\eqref{eq:mean-field}. The position $x$ at which the field is measured would be genuinely arbitrary in the case of (isotropic) single-atom fluorescence.
The mean electric field of an attoquant [equation~\eqref{eq:field-mean-perm}] factorises into an ideally locked temporal structure and the arithmetic mean $\mathbb{E}(\beta)$ of the complex parameters of the constituent coherent states. When many atomic emitters radiate an attoquant, $\mathbb{E}(\beta)$ contains the atomic positions, as already stated after equation~\eqref{eq:Va}. It is this factor alone -- and not the temporal structure, which remains locked regardless -- that renders the radiation process anisotropic.
Where we describe the attoquant state as ``robust'', we mean specifically the temporal (pulse) structure, which is exactly and unconditionally locked; the amplitude, by contrast, is set by $\mathbb{E}(\beta)$ and can be attenuated by the atomic positions in the interaction volume. This geometry-dependent, anisotropic emission of a pulse train with a robustly locked temporal structure in all directions has no classical analogue, and is a distinctive feature of the attoquant description.

Explicitly, the contribution of atom $j$ to harmonic mode $n$ carries a spatial phase-matching factor $\exp\{i[\mathbf{k}_n-(2n+1)\mathbf{k}]\cdot\mathbf{R}_j\}$, where $\mathbf{R}_j$ is the position of the $j$th atom and $\mathbf{k}_n$ the wave vector of the $n$th harmonic mode. Averaged over the cluster, $\mathbb{E}(\beta)$ therefore plays the role of a form factor of the emitter distribution with respect to the wave-vector mismatch $\mathbf{k}_n-(2n+1)\mathbf{k}$. In a similar context, see Ward and New~\cite{PhysRev.185.57}, who performed a detailed classical analysis of the matching problem for high harmonics in the perturbative regime.

This form factor need not vanish in every direction. In fact it remains close to unity within a narrow cone about the phase-matched direction, where $|(\mathbf{k}_n-n\mathbf{k})\cdot\mathbf{R}_j|\ll 1$ for every atom in the cluster. This defines the main propagation direction. Combined with the locked spatiotemporal structure of equation~\eqref{eq:field-mean-perm}, this explains both the directionality and the attosecond time structure of the emitted radiation within the same microscopic picture, and offers a quantum-optical counterpart to the phenomenological phase-matching description familiar from continuous-medium Maxwell theory.
We conjecture that in this picture a more dilute gas or a higher driving frequency -- i.e.\ fewer atoms per coherence volume -- can widen the angular range over which the amplitude remains appreciable.

\subsection{Wider context and outlook}
\label{sec:interference}

The same logic may help to clarify the interpretation of experiments in which interference is observed between harmonics generated at two physically separated locations. Zerne et al. split a single driving beam into two parallel foci under the nozzle of a gas jet~\cite{Zerne1997}, generating two nominally independent harmonic sources. A clean interference pattern was observed in the far field, and interpreted there as evidence that the two sources are locked in phase. In light of the discussion above, such an observation need not imply any pre-existing, externally imposed phase relationship between the two foci: it is exactly the pattern to be expected once the light collected by the detector no longer carries any which-source information, so that the two possible histories -- that the harmonic photon originated from focus $1$ or from focus $2$ -- must be added coherently, as per Heitler's argument.
This distinction between an externally imposed phase relationship and an interference pattern produced by mere indistinguishability is not a subtlety peculiar to HHG. It is precisely what Magyar and Mandel demonstrated for ordinary optical fields, observing clean interference fringes between the light of two completely independent, free-running ruby masers~\cite{MagyarMandel1963}, with no phase relationship whatsoever in any classical sense. Mandel's subsequent quantum-mechanical analysis of that experiment showed that such fringes are the expected signature of the fundamental indistinguishability of the two sources at the level of individual photon counts, and neither require nor imply any stable phase difference between the two independent oscillators~\cite{Mandel1964}.

We stress that the Zerne \emph{et al.} and Magyar-Mandel examples are single-frequency (or narrow-band) illustrations of the general principle~\cite{marcuse1980principles} that indistinguishability, rather than a pre-existing phase relationship, is sufficient for coherent addition. These examples establish that this general principle is not exotic, but are not, and are not intended as, evidence for the specific multi-mode entangled temporal locking claimed for attoquants, which follows instead from the many-atom argument developed above. 
We should also highlight a genuine experimental confound that is easy to conflate with the argument of this section: fluorescence and plasma-recombination radiation, mentioned in section~\ref{sec:intro}, are not obviously subject to the same entanglement mechanism, since they need not originate from the same recollision process at all. Their possible presence is therefore reason for some caution in interpreting any particular interference measurement as a clean, quantitative test of the mechanism proposed here, independently of the conceptual point about indistinguishability.

It is worth placing the coherent permanent state in the context of a broader, and currently active, effort to uncover the genuinely quantum-optical character of high-harmonic generation. The recognition that the emitted light may carry non-trivial quantum correlations initiated the recent quantised-field treatments of single-atom HHG~\cite{GC16,AG19}. At the same time, the entanglement encoded in equation~\eqref{eq:permanent} is one instance of a theme pursued before in a different guise, with the roles of light and matter reversed. In Compton scattering of a photon by a spatially localised electron wave packet, the outgoing photon and the electron end up in a joint state of the form $|\Psi_g(t)\rangle=\int d^2r\,|\Xi(\mathbf r,t)\rangle|\mathbf r\rangle=\sum_{k=-n_0}^{\infty}|\Phi_k(t)\rangle\,|n_0+k\rangle$, such that the photon statistics recorded far from the interaction region depend on where the electron is subsequently found~\cite{VS2008,VS10}. There, it is the electron's position that remains unresolved until measured; here, it is the indistinguishability of the emitter atoms that guarantees the temporal locking and anisotropy, discussed above.

Finally, let us reiterate that the coherent permanent state discussed in this work should not be regarded as a unique construction, but as a representative example of a broader class of symmetric entangled quantum states, naturally appearing in many-atom, indistinguishable-emitter frameworks of HHG~\cite{Varrolphystalk,VarroATTO2019}.
A pure attoquant state is not expected to emerge directly and unconditioned from a standard HHG measurement. In analogy with the high-intensity Schr\"odinger-cat states proposed for HHG, which likewise appear only together with some form of post-selection~\cite{LCPRSKT21}, appropriate conditioning is likely required. Identifying the attoquant state itself is, however, a non-trivial task: the mean field alone cannot distinguish an attoquant from an ordinary mode-locked (separable) state with the same $\mathbb{E}(\beta)$, and demonstrating that the locking specifically originates from entanglement requires a genuinely higher-order signature. The manuscript provides such signatures in the Wigner-function interference structure and the intermodal photon-number correlations and logarithmic negativity (Sec.~3.2). The photon-number distribution in the constituent modes may likewise be a useful indicator. A further such signature follows from filtering out a single harmonic component: its photocount statistics are described by the reduced density operator discussed in the present paper. A concrete detection protocol remains an open question; one plausible, though as yet undeveloped, route would be a conditional measurement correlating the ground-state depletion or ionisation yield across the coherence cluster, read out in post-selection, with the harmonic light collected in a given direction -- similar in spirit to the electron-position conditioning of photon-electron entanglement discussed above~\cite{VS2008}. We regard the identification of concrete, experimentally accessible signatures of this kind as a promising direction opened up by the present work.

\section{Summary}\label{sec:conclusions}

In this work we asked whether attosecond pulse-train formation requires phase-locked harmonics as a matter of physical principle, or only as a consequence of restricting attention to separable quantum states of the driving field. To make this question precise, we generalised the classical notion of pulse synthesis. The semiclassical state and its associated mean-field pulse, as well as the squeezed-vacuum state product and its variance-based pulse, both require an externally imposed mode-locking for highly structured peaks. We traced this requirement to a structural feature of a broad family of separable states that forces the classical mode-locking condition.
This motivated the central construction of the paper: \emph{attoquants}, multimode quantum states whose pulse structure is intrinsically insensitive to such phase-parameter. As a concrete realisation, we introduced the coherent permanent state, a completely symmetric superposition of coherent-state products over all permutations of a fixed displacement-parameter set.

We showed analytically that the mean electric field of the coherent permanent state is a Fourier superposition of the driving harmonics in which every mode shares a single common phase, fixed solely by the arithmetic mean $\mathbb E(\beta)$ of the displacement parameters. As a result, the pulse train remains fully locked for any distribution of the individual phases. This quantum state is genuinely nonclassical in its own right. It is generically entangled between any pair of harmonic modes, with a logarithmic negativity that grows with the spread of $\{\beta_k\}$ parameters, and saturates once interference features in its Wigner function disappear. Its single-mode photon statistics, meanwhile, range from sub- to super-Poissonian depending on the geometry of the parameters. Extending the analysis to the field variance, we found that it is, unlike the mean field, not fully protected against dephasing. Nevertheless, configurations for which the mean-field pulse is suppressed by phase disorder typically show enhanced  values of variance instead.
The coherent permanent state should be understood as one concrete representative of the considerably broader class of attoquants, rather than a unique construction. 

We traced a plausible physical origin for attoquant-like states to the quantum-theoretical indistinguishability of the emitting atoms. Because the coherence volume of the fundamental field is much larger than that of its harmonics, the same cluster of atoms that is driven collectively by the fundamental field radiates its harmonics individually, as spatially resolved sources. Since no observable of the collected light can reveal which atom produced which harmonic photon, the different possible assignments of atoms to modes must be superposed coherently rather than treated as a classical statistical mixture. This construction reproduces exactly the permutation-symmetric structure of the attoquants. It also reframes the question posed at the outset: harmonic phase-locking need not be imposed externally, but can emerge automatically from the indistinguishability of individual emission events.
A pure attoquant state is, however, not expected to emerge unconditioned from a standard HHG measurement. As with the high-intensity HHG-cat states proposed earlier, some form of post-selection is likely required. Testing this proposal is a natural and promising direction for future work.

\ack{
	The ELI ALPS project (GINOP-2.3.6-15-2015-00001) has been supported by the European Union and co-financed by the European Regional Development Fund.
	This research was supported by the National Research, Development and Innovation Office, Hungary ("Quantum Information National Laboratory of Hungary" Grant No. 2022-2.1.1-NL-2022-00004, "Frontline" Research Excellence Programme Grant No. KKP133827).
}
\roles{ Varró Sándor: Conceptualization, Methodology, Funding acquisition, Resources, Supervision, Project administration, Writing — review \& editing. 
	Gombkötő Ákos: Methodology, Investigation, Formal analysis, Software, Data curation, Validation, Visualization, Writing — original draft. }



\appendix



\section{Appendix: Mode-locking}
\setcounter{equation}{0}
\renewcommand{\theequation}{A\arabic{equation}}

\subsection{Mode-locking for semiclassical states}\label{appendixmodelocking}
By Parseval's theorem, $\langle\hat{E}\rangle^2_{\mathrm{SC}\text{(average)}}$ can be expressed as the sum of squares of the Fourier-coefficients in equation~\eqref{classicalEfield}. Consequently, maximising $\eta[\langle\hat{E}\rangle_{\mathrm{SC}}]=\langle\hat{E}\rangle^2_{\mathrm{SC}\text{(peak)}}/\langle\hat{E}\rangle^2_{\mathrm{SC}\text{(average)}}$ at fixed amplitudes $\{|\alpha_n|\}$ reduces to maximising the peak field strength
$\langle\hat{E}\rangle^2_{\mathrm{SC}\text{(peak)}}(\phi_1,\dots\phi_M)
= \max_{x,t}\bigl|\langle\hat{E}(x,t,\phi_1,\dots\phi_M)\rangle \bigr|$,
with respect to the phase parameters $\{\phi_n\}$.
This is a nonlinear optimisation problem in general, but it simplifies considerably by exploiting the bound $\sin\!\big(\omega_n(x/c-t)+\phi_n\big)\leq 1$.
For a chosen spacetime point $(x,t)$, the choice
\begin{equation}\label{eq:mode-locking-phase}
	\phi_n = \frac{\pi}{2} - \omega_n\!\left(\frac{x}{c}-t\right)
	+ 2\pi k,
	\qquad k \in \mathbb{Z},
\end{equation}
causes every term in equation~\eqref{classicalEfield} to attain its maximum simultaneously. Since $(x,t)$ may be chosen arbitrarily, the condition 
is that consecutive phase factors satisfy a uniform difference
$\phi_{n+1} - \phi_n
= -2\omega\!\left(\frac{x}{c}-t\right)
= \text{const}$,
because $\omega_{n+1}-\omega_n = 2\omega$ for all $n$. 
For completeness, the range of achievable values is
\begin{equation}
	\eta[\langle \hat{E} \rangle_\mathrm{SC}]\in\left[1, \tfrac{\left( \sum_{n=1}^M \sqrt{2n_0+2n-1}|\alpha_n| \right)^2}{\sum_{n=1}^M (2n_0+2n-1)|\alpha_n|^2} \right].
\end{equation}

\subsection{Mode-locking for squeezed vacuum states}\label{modelockingsqueezed}

To identify the highest values of $\eta[(\Delta \hat{E})_{\mathrm{SV}}]$, we can use the fact that the time-averaged value of the field-variance is $\mathcal{E}_0^2\sum_{n=1}^M(2n_0+2n-1)(2\sinh^2 r_n+1)$, determined completely by the squeezing amplitudes $\{r_n\}$. 
Consequently, the maximisation problem is identical with the maximisation of the peak field variance $(\Delta\hat{E})^2_{\mathrm{SV}\text{(peak)}}(\theta_1,\dots \theta_M) = \max_{x,t}\big(\Delta\hat E\big)^2_{SV}(x,t,\theta_1,\dots \theta_M)$, at fixed $\{r_n\}$ squeezing amplitudes.
Analogously to the above, it is maximised when each cosine term in equation~\eqref{eq:squeezed-mode-variance} reaches its maximum simultaneously. For a chosen  $(x,t)$, the choice 
\begin{equation}
	\theta_n = -2\omega_n\Big(\frac{x}{c}-t\Big) + 2\pi k, \qquad k\in\mathbb Z,
	\label{eq:variance-locking-phase}
\end{equation}
does exactly that. Substituting $\omega_{n+1}-\omega_n=2\omega$ into equation~\eqref{eq:variance-locking-phase}, the difference of consecutive squeezing phase factors reads
$\theta_{n+1}-\theta_n = -4\omega(\tfrac{x}{c}-t) = \mathrm{const}$.
The achievable range of $\eta[(\Delta \hat{E})_\mathrm{SV}]$ is
\begin{equation}
	\eta[(\Delta \hat{E})_\mathrm{SV}]\in\left[1, \tfrac{\sum_{n=1}^M(2n_0+2n-1)\left(2\sinh^2 r_n+1+\sinh(2r_n)\right)}{\sum_{n=1}^M(2n_0+2n-1)(2\sinh^2 r_n+1)} \right].
\end{equation}

\subsection{Field-cumulants and mode-locking in separable states}\label{cumulantappendix}
The statistics of $\hat E(x,t)=\sum_{n=1}^M\hat E_n(x,t)$, at a fixed spacetime point, are captured by the characteristic function $\chi(\lambda;x,t)=\langle\Psi|e^{i\lambda\hat E(x,t)}|\Psi\rangle$, whose cumulants $\langle\kappa_j[\hat E(x,t)]\rangle$ are defined through
\begin{equation}
	\ln\chi(\lambda;x,t) = \sum_{j=1}^{\infty}\frac{(i\lambda)^j}{j!}\langle\kappa_j[\hat E(x,t)]\rangle.
	\label{eq:cumulant-gendef}
\end{equation}
Accordingly, we define the \emph{field $j$th-cumulant pulse synthesis} as the pursuit of a high value of $\eta[\kappa_j[\hat E]]$. Note that $\kappa_1,\kappa_2,\kappa_3$ coincide with $\hat E$, $(\hat E-\langle\hat E\rangle)^2$, and $(\hat E-\langle\hat E\rangle)^3$, respectively. From $j=4$ onward, $\kappa_j$ is a fixed polynomial combination of moments (e.g. $\kappa_4=\langle(\Delta\hat E)^4\rangle-3\langle(\Delta\hat E)^2\rangle^2$). The structural result below is insensitive to this distinction.
In general, due to $[\hat E_n,\hat E_m]=0$ for $n\neq m$, the factorisation $e^{i\lambda\hat E}=\prod_n e^{i\lambda\hat E_n}$ holds. For separable $|\Psi\rangle=\bigotimes_{n=1}^M|\psi_n\rangle_n$ states, the expectation value of this exponential factorises as well, resulting in
\begin{equation}
	\chi(\lambda;x,t) = \prod_{n=1}^{M}\chi_n(\lambda;x,t), \qquad \chi_n(\lambda;x,t)\equiv\langle\psi_n|e^{i\lambda\hat E_n(x,t)}|\psi_n\rangle_n.
	\label{eq:cumulant-additivity}
\end{equation}
Taking the logarithm and collecting terms by powers of $\lambda$ yields 
\begin{equation}
	\langle\kappa_j[\hat E(x,t)]\rangle = \sum_{n=1}^{M}\langle\kappa_j[\hat E_n(x,t)]\rangle \qquad\text{for every } j\geq1.
	\label{eq:cumulant-additivity-2}
\end{equation}
Additivity of the cumulants holds for any separable product state. Note that additivity at $j=1$ is in fact unconditional, following from linearity of the expectation value alone, $\langle\hat E\rangle=\sum_n\langle\hat E_n\rangle$, regardless of whether $|\Psi\rangle$ is separable.


Provided that the quantum state and the cumulants are mathematically well-behaved, the terms in the $j$th field cumulant can quite generally be written as a Fourier-series
\begin{equation}\label{eq:Nth-order-ansatz}
	\kappa_j\bigl[\hat E_n(x,t)\bigr] \;=\; a_{j,n}(\{\Lambda_{k,n}\}) \;+\;
	\sum_{p=1}^{} b_{j,n,p}(\{\Lambda_{k,n}\})\cos\Bigl(p\omega_n\bigl(\tfrac{x}{c}-t\bigr) +
	f_{j,n,p}\left(\{\varphi_{k,n}\}\right)\Bigr),
\end{equation}
where $a_{j,n}(\cdot)$, $b_{j,n,p}(\cdot)$, and $f_{j,n,p}(\cdot)$ are functions of the amplitude and phase-parameter.
By equation~\eqref{eq:cumulant-additivity-2}, the cumulants can also be written as a Fourier-series, in analogy with equation~\eqref{eq:Nth-order-ansatz}. Collecting terms by frequency $\omega'=p\omega_n$, one can determine the effective phases of the cumulant's Fourier components as 
\begin{equation}\label{cumulantphase}
	\text{atan2}\left(
	\sum_{\substack{n,p\\p\omega_n=\omega'}}
	b_{j,n,p}(\{\Lambda_{k,n}\})\sin\Bigl(
	f_{j,n,p}\left(\{\varphi_{k,n}\}\right)\Bigr)
	,
	\sum_{\substack{n,p\\p\omega_n=\omega'}}
	b_{j,n,p}(\{\Lambda_{k,n}\})\cos\Bigl(
	f_{j,n,p}\left(\{\varphi_{k,n}\}\right)\Bigr)
	\right) .
\end{equation}

\section{Appendix: Density matrix and related quantities of the coherent permanent state} 
\setcounter{equation}{0}
\renewcommand{\theequation}{B\arabic{equation}}

\subsection{Fock-basis coefficients}\label{appendixfockcoeff}

Straightforward calculation gives the Fock-basis coefficients $C_{q_1,\ldots,q_M}(t)$ as
\begin{eqnarray}
	C_{q_1,\ldots,q_M}(t) &=& \langle q_1,\ldots,q_M|\Psi(t)\rangle_{\rm PERM} = \frac{1}{\sqrt{\mathcal N}}\sum_{\sigma\in S_M}\prod_{n=1}^{M}e^{-|\beta_{\sigma(n)}|^2/2}\frac{\beta_{\sigma(n)}^{q_n}\,e^{-iq_n\omega_n t}}{\sqrt{q_n!}} \nonumber\\
	&=& \frac{1}{\sqrt{\mathcal N}}\,\frac{e^{-i\sum_n q_n\omega_n t}}{\prod_n\sqrt{q_n!}}\,e^{-\frac12\sum_k|\beta_k|^2}\,\mathrm{perm}\big(\mathbf{B}_{q_1,\ldots,q_M}\big),
	\label{eq:photon-dist}
\end{eqnarray}
where the factor $e^{-\frac{1}{2}\sum_k|\beta_k|^2}$ is, of course, independent
of the permutation $\sigma$, and $\mathbf{B}_{q_1,\ldots,q_M}\in\mathbb{C}^{M\times M}$
is the matrix defined elementwise as $\bigl(\mathbf{B}_{q_1,\ldots,q_M}\bigr)_{n,k} = \beta_k^{q_n}$, explicitly
\begin{equation}
	\mathbf{B}_{q_1,\ldots,q_M}
	=
	\begin{pmatrix}
		\beta_1^{q_1} & \beta_2^{q_1} & \cdots & \beta_M^{q_1} \\
		\vdots        & \vdots        & \ddots & \vdots        \\
		\beta_1^{q_M} & \beta_2^{q_M} & \cdots & \beta_M^{q_M}
	\end{pmatrix}.
\end{equation}

\subsection{One-mode reduced density matrix}\label{densityM}
Here we outline the steps involved in partially tracing the density matrix of $|\Psi(t)\rangle_{\rm PERM}$ over the overcomplete coherent-state basis. We apply the resolution of the identity
$\mathbf{1} = \frac{1}{\pi}\int_{\mathbb C}|\alpha\rangle\langle\alpha|\,\mathrm d^2\alpha$,
with which partial tracing over the $j$th mode can be carried out as
\begin{equation}
	\mathrm{Tr}_j(\rho) = \frac{1}{\pi}\int_{\mathbb C}\,{}_j\!\langle\alpha|\rho|\alpha\rangle_j\,\mathrm d^2\alpha.
	\label{eq:partial-trace}
\end{equation}
For a one-mode density matrix $\rho=|\beta'\rangle\langle\beta''|$, this gives $\mathrm{Tr}(\rho) = \tfrac{1}{\pi}\int_{\mathbb{C}}\langle\alpha|\beta'\rangle\langle\beta''|\alpha\rangle\,\mathrm d^2\alpha = \langle\beta''|\beta'\rangle$. Using this result, tracing $\rho$ over every mode except $N$ replaces each factor 
$|\beta_{\sigma(n)}\rangle_n\langle\beta_{\sigma'(n)}|_n$, $n\ne N$, by the scalar $\langle\beta_{\sigma'(n)}|\beta_{\sigma(n)}\rangle=K_{\sigma'(n),\sigma(n)}$.
Carrying out the same integration $M-1$ times results in 
\begin{equation}\label{reducingdensmat}
	\rho_N \;=\; \frac{1}{\mathcal N}\sum_{\sigma,\sigma'\in S_M}\left(\prod_{n\neq N} K_{\sigma'(n),\sigma(n)}\right)|\beta_{\sigma(N)}\rangle\langle\beta_{\sigma'(N)}|.
\end{equation}
Set $j=\sigma(N)$ and $k=\sigma'(N)$. Restricting $\sigma$ and $\sigma'$ to the domain $\overline N=\{1,\dots,M\}\setminus\{N\}$ then defines bijections $\widetilde\sigma$ and $\widetilde\sigma'$ on $\overline N$ whose ranges exclude $j$ and $k$, respectively. Thus, we can collect the terms by $j$ and $k$ as
\begin{equation}
	\rho_N=\frac1{\mathcal N}\sum_{j,k=1}^M|\beta_j\rangle\langle\beta_k|
	\sum_{\widetilde\sigma} \sum_{\widetilde\sigma'}
	\prod_{n\in\overline N}K_{\widetilde\sigma'(n),\widetilde\sigma(n)}.
\end{equation}
For fixed $\widetilde\sigma$, substituting $\pi=\widetilde\sigma'\circ\widetilde\sigma^{-1}$ and $m=\widetilde\sigma(n)$ into the inner sum results in
\begin{equation}
	\sum_{\widetilde\sigma'}\prod_{n\in\overline N}K_{\widetilde\sigma'(n),\widetilde\sigma(n)}
	=\sum_{\pi}\prod_{m\neq j} K_{\pi(m),m}=\mathrm{perm}\bigl(\mathbf{\Omega}_{k,j}\bigr),
\end{equation}
where $\mathbf \Omega_{k,j}$ is the $(M{-}1)\times(M{-}1)$ submatrix of $\mathbf K$ with row $k$ and column $j$ deleted. This term is
independent of $\widetilde\sigma$. Summing over the $(M-1)!$ choices of $\widetilde\sigma$ gives $(M-1)!\,\mathrm{perm}(\mathbf{\Omega}_{k,j})$. Therefore
\begin{equation}\label{eq:rhoN}
	\rho_N=\frac{(M-1)!}{\mathcal N}\sum_{j,k=1}^M\mathrm{perm}\bigl( \mathbf{\Omega}_{k,j} \bigr)\,|\beta_j\rangle\langle\beta_k|.
\end{equation}
By mode-permutation symmetry, this is independent of the mode $N$.

\subsection{Two-mode reduced density matrix}\label{densityNR}

An identical argument can be followed as in Appendix~\ref{densityM}, up to equation~\eqref{reducingdensmat}. Tracing out all modes except those indexed by $N$ and $R$ restricts the permutations to the domain $\overline{N,R}=\{1,\dots,M\}\setminus\{N,R\}$. Writing $j=\sigma(N)$, $p=\sigma(R)$, $k=\sigma'(N)$, $q=\sigma'(R)$ -- necessarily $j\neq p$ and $k\neq q$, since $\sigma,\sigma'$ are injective -- the restricted maps $\widetilde\sigma$ and $\widetilde\sigma'$ act on the common domain $\overline{N,R}$, with ranges excluding $\{j,p\}$ and $\{k,q\}$, respectively, giving
\begin{equation}
	\rho_{NR}=\frac1{\mathcal N}\sum_{\substack{j\ne p\\k\ne q}}|\beta_j\rangle\langle\beta_k|_N\otimes|\beta_p\rangle\langle\beta_q|_R\sum_{\widetilde\sigma}\sum_{\widetilde\sigma'} \prod_{n\in\overline{N,R}}K_{\widetilde\sigma'(n),\widetilde\sigma(n)}.
\end{equation}
The inner double sum evaluates to $(M-2)!\,\mathrm{perm}(\mathbf \Omega_{kq,jp})$, independent of $\widetilde\sigma$. Hence
\begin{equation}\label{eq:rhoNR}
	\rho_{NR}=\frac{(M-2)!}{\mathcal N}\sum_{\substack{j,p=1\\ j\neq p}}^{M}\ \sum_{\substack{k,q=1\\ k\neq q}}^{M} \mathrm{perm}\bigl(\mathbf \Omega_{kq,jp}\bigr)\,|\beta_j\rangle\langle\beta_k|_N\otimes|\beta_p\rangle\langle\beta_q|_R,
\end{equation}
with $\mathbf \Omega_{kq,jp}$ the $(M{-}2)\times(M{-}2)$ submatrix of $\mathbf K$ with rows $k,q$ and columns $j,p$ deleted. The reduced density matrix is independent of $\{N,R\}$.

\subsection{Wigner function}\label{wignerappendix}
For completeness, we recall that the (symmetrically ordered) characteristic function of an operator $\hat A$ and its associated Wigner function are given by
\begin{equation}
	\chi_A(\lambda) = \mathrm{Tr}\bigl[\hat A\,\hat D(\lambda)\bigr], \qquad \hat D(\lambda) = e^{\lambda \hat a^\dagger - \lambda^* \hat a}, \qquad W_A(\alpha) = \frac{1}{\pi^2}\int d^2\lambda\; \chi_A(\lambda)\,e^{\alpha\lambda^*-\alpha^*\lambda}.
\end{equation}
Since $\rho_N$ in equation~\eqref{eq:rhoN} is a sum of dyads $|\beta_j\rangle\langle\beta_k|$, we first determine the Wigner function of these terms. Using $\hat D(\lambda)|\beta_j\rangle = e^{(\lambda\beta_j^*-\lambda^*\beta_j)/2}|\beta_j+\lambda\rangle$,
\begin{equation}
	\chi_{|\beta_j\rangle\langle\beta_k|}(\lambda) = \langle\beta_k|\hat D(\lambda)|\beta_j\rangle = e^{(\lambda\beta_j^*-\lambda^*\beta_j)/2}\,\langle\beta_k|\beta_j+\lambda\rangle.
\end{equation}
Expanding $\langle\beta_k|\beta_j+\lambda\rangle=\exp\!\left(-\tfrac12|\beta_k|^2-\tfrac12|\beta_j+\lambda|^2+\beta_k^*(\beta_j+\lambda)\right)$ and simplifying leads to
\begin{equation}
	\chi_{|\beta_j\rangle\langle\beta_k|}(\lambda) = K_{kj}\,\exp\!\left(-\tfrac12|\lambda|^2+\lambda\beta_k^*-\lambda^*\beta_j\right).
\end{equation}
Substituting into the Wigner function and using the integral relation $\int d^2\lambda\, e^{-c|\lambda|^2+\lambda A-\lambda^*B}=\frac{\pi}{c}e^{-AB/c}$ with $c=\tfrac12$, $A=\beta_k^*-\alpha^*$, and $B=\beta_j-\alpha$, the result can be given in closed form
\begin{equation}\label{eq:Wdyad}
	W_{|\beta_j\rangle\langle\beta_k|}(\alpha) \;=\; \frac{2}{\pi}\,K_{kj}\,\exp\!\left[-2(\alpha-\beta_j)(\alpha^*-\beta_k^*)\right].
\end{equation}
By linearity of the Wigner transform, applying equation~\eqref{eq:Wdyad} term by term to equation~\eqref{eq:rhoN} results in
\begin{equation}\label{eq:WrhoN}
	W_{\rho_N}(\alpha) \;=\; \frac{2\,(M-1)!}{\pi\,\mathcal N}\sum_{j,k=1}^{M} K_{kj}\;\mathrm{perm}\bigl(\mathbf \Omega_{k,j}\bigr)\,\exp\!\left[-2(\alpha-\beta_j)(\alpha^*-\beta_k^*)\right],
\end{equation}
where $\alpha=x+iy$ is the phase-space variable.

\subsection{Logarithmic negativity}\label{applogneg}

The logarithmic negativity is defined as $E_N(\rho_{NR})=\log_2\|\rho_{NR}^{T_R}\|_1$, where $T_R$ is the partial transpose on mode $R$ and $\|\cdot\|_1$ the trace norm.
The partial transpose of a coherent-state dyad can be written simply as 
\begin{equation}\label{eq:PTrule}
	T\bigl(|\beta\rangle\langle\beta'|\bigr)=|\beta'^{*}\rangle\langle\beta^{*}|.
\end{equation}
Applying \eqref{eq:PTrule} to the $R$-factor of each term in equation~\eqref{eq:rhoNR},
\begin{equation}\label{eq:rhoNRPT}
	\rho_{NR}^{T_R}=\frac{(M-2)!}{\mathcal N}\sum_{\substack{j\ne p\\k\ne q}}\mathrm{perm}\bigl(\mathbf \Omega_{kq,jp}\bigr)\,|\beta_j\rangle\langle\beta_k|_N\otimes|\beta_q^{*}\rangle\langle\beta_p^{*}|_R.
\end{equation}
Write $\mathbf K=V^\dagger V$ for an invertible $V\in\mathbb C^{M\times M}$, and introduce two auxiliary orthonormal bases $\{|e_l\rangle\}_{l=1}^M$, $\{|f_m\rangle\}_{m=1}^M$ (spanning the relevant subspaces of modes $N$ and $R$, respectively) via
$|\beta_j\rangle_N=\sum_l V_{lj}|e_l\rangle$, $|\beta_p^{*}\rangle_R=\sum_m V^*_{mp}\,|f_m\rangle$. In this basis, equation~\eqref{eq:rhoNRPT} takes the matrix form
$\rho_{NR}^{T_R}\ =\ \frac{(M-2)!}{\mathcal N}\,(V\otimes\overline V)\,\mathbf T\,(V\otimes\overline V)^\dagger$,
where $\mathbf T$ is the $M^2\times M^2$ matrix, with rows and columns labelled by ordered pairs, whose $\bigl((j,q),(k,p)\bigr)$ entry is
\begin{equation}
	T_{(j,q),(k,p)}=\mathrm{perm}\bigl(\mathbf \Omega_{kq,jp}\bigr)\,(1-\delta_{jp})(1-\delta_{kq}).
\end{equation}
Using the eigenvalues $\{\lambda_i\}$ of $(\mathbf K\otimes\mathbf K^{*})\mathbf T$, we can express the logarithmic negativity as
\begin{equation}\label{eq:ENfinal}
	E_N\left[\rho_{NR}\right] \;=\; \log_2\!\left[\frac{(M-2)!}{\mathcal N}\sum_i\bigl|\lambda_i\bigr|\right].
\end{equation}

\section{Appendix: Expectation values in the coherent permanent state}
\setcounter{equation}{0}
\renewcommand{\theequation}{C\arabic{equation}}

\subsection{One-mode expectation values}\label{sec:one-mode}
From the definition of $|\Psi(t)\rangle_{\mathrm{PERM}}$, together with the coherent-state eigenvalue relation $\hat{a}_n|\beta_{\sigma'(n)}\,e^{-i\omega_n t}\rangle_n = \beta_{\sigma'(n)}\,e^{-i\omega_n t}|\beta_{\sigma'(n)}\,e^{-i\omega_n t}\rangle_n$, we obtain
\begin{equation}
	\Bigl\langle \hat{a}^{\dagger k}_n \hat{a}^l_n \Bigr\rangle
	=
	\frac{1}{\mathcal{N}}
	\sum_{\sigma,\sigma'\in S_M}
	\beta^{*k}_{\sigma(n)}\,\beta^{l}_{\sigma'(n)}\,
	e^{i(k-l)\omega_n t}
	\prod_{m=1}^{M} K_{\sigma(m),\sigma'(m)}.
\end{equation}
Setting $p=\sigma(m)$, $p'=\sigma(n)$ and correspondingly $m=\sigma^{-1}(p)$, $n=\sigma^{-1}(p')$ together with $\tau=\sigma'\circ\sigma^{-1}$,
the product becomes $\prod_p K_{p,\tau(p)}$. As $\sigma$ ranges over
$S_M$ for fixed $\tau$, the image $p'=\sigma(n)$ takes each of $1,\ldots,M$ values exactly $(M-1)!$ times.  Hence
\begin{align}
	\Bigl\langle \hat{a}^{\dagger k}_n \hat{a}^l_n \Bigr\rangle
	=
	\frac{1}{\mathcal{N}}
	\sum_{\sigma,\tau\in S_M}
	\beta^{*k}_{p'}\, \beta^{l}_{\tau(p')}\,e^{i(k-l)\omega_n t} 
	\prod_{p=1}^{M} K_{p,\tau(p)} 
	=
	\frac{(M-1)!}{\mathcal{N}}\,e^{i(k-l)\omega_n t}
	\sum_{j=1}^{M}\beta^{*k}_j
	\sum_{\tau\in S_M}\beta^{l}_{\tau(j)}
	\prod_{p=1}^M K_{p,\tau(p)}.
\end{align}
The inner $\tau$-sum is the permanent of the matrix
$\mathbf{K}^{[j]}_l$, defined as $\mathbf{K}$ with row $j$ replaced
element-wise by $\bigl(\beta_v^l\,K_{j,v}\bigr)_v$, so that
$ \sum_{\tau\in S_M}\beta^{l}_{\tau(j)}\prod_{p=1}^M K_{p,\tau(p)}
= \mathrm{perm}\bigl(\mathbf{K}^{[j]}_l\bigr)$.
Using $\mathcal N=M!\,\mathrm{perm}(\mathbf K)$,
\begin{equation}
	\Bigl\langle \hat{a}^{\dagger k}_n \hat{a}^l_n \Bigr\rangle
	=
	\frac{e^{i(k-l)\omega_n t}}{M\cdot\mathrm{perm}(\mathbf{K})}
	\sum_{j=1}^{M} \beta^{*k}_j \mathrm{perm}\bigl(\mathbf{K}^{[j]}_{l}\bigr),
\end{equation}
where $\mathbf{K}^{[j]}_{l}$ denotes the $\mathbf{K}$ matrix with row $j$
replaced element-wise by $(\beta_1^l K_{j,1}, \dots \beta_M^l K_{j,M})$. 

\subsection{Two-mode expectation values}\label{twomodeexpectation}
We calculate the general two-mode expectation value
$\langle\hat{a}^{\dagger k}_n\hat{a}^l_n\hat{a}^{\dagger f}_m\hat{a}^g_m\rangle$
for distinct modes $n\neq m$.  By the same procedure as above,
\begin{align}\label{twomodefull}
	\Bigl\langle\hat{a}^{\dagger k}_n\hat{a}^l_n
	\hat{a}^{\dagger f}_m\hat{a}^g_m\Bigr\rangle
	&=
	\frac{1}{\mathcal{N}}
	\sum_{\sigma,\sigma'\in S_M}
	\beta^{*k}_{\sigma(n)}\,\beta^l_{\sigma'(n)}\,
	\beta^{*f}_{\sigma(m)}\,\beta^g_{\sigma'(m)}\,
	e^{i[(k-l)\omega_n+(f-g)\omega_m]t}
	\prod_{j=1}^M K_{\sigma(j),\sigma'(j)}.
\end{align}
After re-indexing with $\tau=\sigma'\circ\sigma^{-1}$ and setting
$p'=\sigma(n)$, $p''=\sigma(m)$, the ordered pair $(p',p'')=(\sigma(n),\sigma(m))$ visits every
ordered pair of distinct indices $(j,j')$ ($j\neq j'$) exactly
$(M-2)!$ times as $\sigma$ ranges over $S_M$, so that
\begin{equation}
	\sum_{\sigma\in S_M}\beta_{p'}^{*k}\,\beta_{\tau(p')}^{l}\,\beta_{p''}^{*f}\,\beta_{\tau(p'')}^{g} = (M-2)!\sum_{\substack{j,j'=1\\ j\neq j'}}^{M}\beta_j^{*k}\,\beta_{\tau(j)}^{l}\,\beta_{j'}^{*f}\,\beta_{\tau(j')}^{g}.
	\label{eq:two-mode-reindexed}
\end{equation}
Defining
$\mathbf{K}^{[j,j']}_{l,g}$ as $\mathbf{K}$ with row $j$ replaced
by $\bigl(\beta^l_v K_{j,v}\bigr)_v$ and row $j'$ replaced by
$\bigl(\beta^g_v K_{j',v}\bigr)_v$, we obtain
\begin{equation}
	\Bigl\langle\hat{a}^{\dagger k}_n\hat{a}^l_n
	\hat{a}^{\dagger f}_m\hat{a}^g_m\Bigr\rangle
	=
	\frac{e^{i[(k-l)\omega_n+(f-g)\omega_m]t}}
	{M(M-1)\,\mathrm{perm}(\mathbf{K})}
	\sum_{\substack{j,j'=1\\j\neq j'}}^{M}
	\beta^{*k}_j\,\beta^{*f}_{j'}\,
	\mathrm{perm}\bigl(\mathbf{K}^{[j,j']}_{l,g}\bigr).
\end{equation}
The results simplify considerably for $l=g=0$ or $k=f=0$. For $k=f=0$, $\mathbf{K}^{[j,j']}_{l,g}$ carries no $\beta^*$ weight on rows $j,j'$, so that
\begin{equation}
	\big\langle\hat a_n^l\hat a_m^g\big\rangle = \frac{e^{-i(l\omega_n+g\omega_m)t}}{M(M-1)}\sum_{\substack{j,j'=1\\j\neq j'}}^{M}\beta_j^{l}\,\beta_{j'}^{g}.
\end{equation}
For any bijection $\tau$, we can write 
$\sum_{j,j'=1,~j\neq j'}^{M}\beta_{\tau(j)}^{l}\,\beta_{\tau(j')}^{g} = \Big(\sum_j \beta_{\tau(j)}^l\Big)\Big(\sum_{j'}\beta_{\tau(j')}^g\Big) - \sum_j\beta_{\tau(j)}^{l+g}$ $= M^2 \mathbb E(\beta^l)\mathbb E(\beta^g) - M\,\mathbb E(\beta^{l+g})$,
which, applied with $\tau=\mathrm{id}$, gives
$\big\langle\hat a_n^l\hat a_m^g\big\rangle = \frac{e^{-i(l\omega_n+g\omega_m)t}}{M-1}\Big[M\,\mathbb{E}(\beta^l)\mathbb{E}(\beta^g) - \mathbb{E}(\beta^{l+g})\Big]$,
and by Hermitian conjugation,
$\big\langle\hat a_n^{\dagger k}\hat a_m^{\dagger f}\big\rangle = \frac{e^{i(k\omega_n+f\omega_m)t}}{M-1}\Big[M\,\mathbb{E}(\beta^{*k})\mathbb{E}(\beta^{*f}) - \mathbb{E}(\beta^{*(k+f)})\Big]$.

\subsection{First-order mixed intermodal term}\label{mixedcorr}

By equation~\eqref{twomode} with $k=1,l=0,f=0,g=1$,
\begin{equation}\label{eq:mixed-start}
	\langle\hat a_n^\dagger\hat a_m\rangle = \frac{e^{i(\omega_n-\omega_m)t}}{M(M-1)\,\mathrm{perm}(\mathbf{K})}\sum_{\substack{j,j'=1\\j\neq j'}}^M\beta^*_j\,\mathrm{perm}\bigl(\mathbf{K}^{[j,j']}_{0,1}\bigr), \qquad n\neq m.
\end{equation}
Setting $l=0$ leaves row $j$ of $\mathbf{K}^{[j,j']}_{0,1}$ equal to row $j$ of $\mathbf K$ itself (since $\beta_v^0=1$), while row $j'$ is weighted by $\beta_v$, hence $\mathrm{perm}\bigl(\mathbf{K}^{[j,j']}_{0,1}\bigr)=\mathrm{perm}\bigl(\mathbf{K}^{[j']}_{1}\bigr)$, independent of $j$, and the sum in equation~\eqref{eq:mixed-start} becomes
$ S\equiv \sum_{\substack{j,j'=1\\j\neq j'}}^M\beta^*_j\,\mathrm{perm}\bigl(\mathbf{K}^{[j']}_{1}\bigr)$.
Two consequences of results already established are needed to evaluate $S$ in closed form. The first, $\sum_{j'=1}^M\beta^*_{j'}\,\mathrm{perm}\bigl(\mathbf{K}^{[j']}_1\bigr)=M\,\mathrm{perm}(\mathbf{K})\,\langle\hat N_n\rangle$, follows from equation~\eqref{eq:photonmean} (with $j\to j'$). The second follows the same way from the $k=0$ case of equation~\eqref{adagkal}: setting $k=0$ gives $\langle\hat a_n^l\rangle=\frac{e^{-il\omega_nt}}{M\,\mathrm{perm}(\mathbf K)}\sum_j\mathrm{perm}\bigl(\mathbf K^{[j]}_l\bigr)$, which, compared with the closed form $\langle\hat a_n^l\rangle=\mathbb E(\beta^l)e^{-il\omega_nt}$ of equation~\eqref{adagkexp}, gives
$\sum_{j=1}^M\mathrm{perm}\bigl(\mathbf K^{[j]}_l\bigr) = M\,\mathrm{perm}(\mathbf K)\,\mathbb E(\beta^l)$.
Since $\mathrm{perm}\bigl(\mathbf K^{[j']}_1\bigr)$ does not depend on $j$, the above-mentioned sum over $j\neq j'$ results in
\begin{equation}
	S = \Bigl(\sum_{j=1}^M\beta_j^*\Bigr)\Bigl(\sum_{j'=1}^M\mathrm{perm}\bigl(\mathbf K^{[j']}_1\bigr)\Bigr) - \sum_{j=1}^M\beta_j^*\,\mathrm{perm}\bigl(\mathbf K^{[j]}_1\bigr).
\end{equation}
The two formulae given above lead to
$S = M \mathbb E(\beta^*)\cdot M\,\mathrm{perm}(\mathbf K)\,\mathbb E(\beta) - M\,\mathrm{perm}(\mathbf K)\,\langle\hat N_n\rangle $ $= M\,\mathrm{perm}(\mathbf K)\Bigl[M\,|\mathbb E(\beta)|^2-\langle\hat N_n\rangle\Bigr]$.
Inserting this into equation~\eqref{eq:mixed-start} gives the closed form
\begin{equation}
	\langle\hat a_n^\dagger\hat a_m\rangle = \frac{e^{i(\omega_n-\omega_m)t}}{M-1} \left[  M|\mathbb{E}(\beta)|^2-\langle\hat N_n\rangle \right] \qquad n\neq m,
\end{equation}
i.e.\ calculating the first-order intermodal correlator requires no permanent evaluation beyond that already needed for the mean photon number $\langle\hat N_n\rangle$.

\subsection{Non-negativity of photon-number deviation}\label{appDeltaN}

The quantity $\langle\delta\hat{N}\rangle\equiv\langle\hat N_n\rangle-|\mathbb E(\beta)|^2$, used throughout section~\ref{attoquant}, is non-negative. For mode $n$, define operator $\delta\hat a_n \equiv \hat a_n - \langle\hat a_n\rangle$. Expanding its normally-ordered second moment,
\begin{equation}
	\langle\delta\hat a_n^\dagger\,\delta\hat a_n\rangle
	= \langle\hat a_n^\dagger\hat a_n\rangle - \langle\hat a_n\rangle\langle\hat a_n^\dagger\rangle - \langle\hat a_n^\dagger\rangle\langle\hat a_n\rangle + \langle\hat a_n^\dagger\rangle\langle\hat a_n\rangle
	= \langle\hat N_n\rangle - |\langle\hat a_n\rangle|^2.
\end{equation}
For the coherent permanent state, $\langle\hat a_n\rangle=\mathbb E(\beta)\,e^{-i\omega_nt}$, so
$\langle\delta\hat{N}\rangle = \langle\hat N_n\rangle-|\mathbb E(\beta)|^2 = \langle\delta\hat a_n^\dagger\,\delta\hat a_n\rangle$.
The right-hand side is the squared norm of a state vector,
\begin{equation}
	\langle\delta\hat a_n^\dagger\,\delta\hat a_n\rangle = \bigl\|\delta\hat a_n\,|\Psi(t)\rangle\bigr\|^2 \;\geq\;0,
\end{equation}
non-negative for \emph{any} $|\Psi(t)\rangle$. Hence  $\langle\delta\hat{N}\rangle\geq0 $
for every coherent permanent state.

A related, but stronger claim is used with respect to equation~\eqref{eq:total-variance-perm}: that the whole term proportional to $\langle\delta\hat N\rangle$ in the total field variance is non-negative.
This term can be written succinctly as $\frac{2\mathcal E_0^2}{M-1}\langle\delta\hat N\rangle\bigl[M^2(M+2n_0)-|\mathcal C_1(\tau)|^2\bigr]$, where $w_n\equiv2n_0+2n-1$,
$\mathcal C_1(\tau)\equiv\sum_{n=1}^M\sqrt{w_n}\,e^{i\omega_n\tau}$, and we use $W\equiv\sum_{n=1}^Mw_n=M(M+2n_0)$.
Since $\langle\delta\hat N\rangle\geq0$ has just been shown, it remains to show that the bracket, $MW-|\mathcal C_1(\tau)|^2$, is non-negative. Applying the Cauchy--Schwarz inequality to the vectors $u_n=\sqrt{w_n}\,e^{i\omega_n\tau}$ and $v_n=1$,
\begin{equation}
	\bigl|\mathcal C_1(\tau)\bigr|^2 = \Bigl|\sum_{n=1}^M u_nv_n\Bigr|^2 \leq \Bigl(\sum_{n=1}^M|u_n|^2\Bigr)\Bigl(\sum_{n=1}^M|v_n|^2\Bigr) = W\cdot M = M^2(M+2n_0),
\end{equation}
so the first quantum contribution to equation~\eqref{eq:total-variance-perm} is non-negative for every coherent permanent state and every $(x,t)$.

\subsection{Summation in the variance}\label{sumvari}

This appendix derives the total field-variance formula, equation~\eqref{eq:total-variance-perm}, by summing the single-mode variances, equation~\eqref{eq:single-mode-variance-full}, and the covariances, equation~\eqref{eq:field-cov-closed}, over all $M$ modes. With $w_n\equiv2n_0+2n-1$, $W\equiv\sum_nw_n=M(M+2n_0)$, and $\mathcal C_1(\tau)$ as in Appendix~\ref{appDeltaN}. Introduce also
$\mathcal C_2(\tau)\equiv\sum_{n=1}^M w_n\,e^{2i\omega_n\tau}$, with which one can write 
$\mathcal C_1(\tau)^2=\sum_{n,m}\sqrt{w_nw_m}\,e^{i(\omega_n+\omega_m)\tau}=\mathcal C_2(\tau)+\sum_{n\neq m}\sqrt{w_nw_m}\,e^{i(\omega_n+\omega_m)\tau}$; likewise $|\mathcal C_1(\tau)|^2=\sum_{n,m}\sqrt{w_nw_m}\,e^{i(\omega_n-\omega_m)\tau}=W+\sum_{n\neq m}\sqrt{w_nw_m}\,e^{i(\omega_n-\omega_m)\tau}$. Hence, we get
\begin{equation}\label{eq:sumvari-identities}
	\sum_{n\neq m}\sqrt{w_nw_m}\,e^{i(\omega_n+\omega_m)\tau}=\mathcal C_1(\tau)^2-\mathcal C_2(\tau), \qquad
	\sum_{n\neq m}\sqrt{w_nw_m}\cos\bigl[(\omega_n-\omega_m)\tau\bigr]=|\mathcal C_1(\tau)|^2-W.
\end{equation}
Summing equation~\eqref{eq:single-mode-variance-full} over $n$ gives the single-mode contribution,
\begin{equation}
	\sum_{n=1}^M\bigl(\Delta\hat E_n(\tau)\bigr)^2_{\rm PERM} = \mathcal E_0^2\Bigl[(1+2\langle\delta\hat N\rangle)\,W - 2\,\mathrm{Re}\bigl(\mathrm{Var}_{\mathbb C}(\beta)\,\mathcal C_2(\tau)\bigr)\Bigr].
\end{equation}
Summing equation~\eqref{eq:field-cov-closed} over $n\neq m$ and inserting equation~\eqref{eq:sumvari-identities}, yields
\begin{align}
	\sum_{n\neq m}\mathrm{Cov}(\hat E_n,\hat E_m) &= \frac{2\mathcal E_0^2}{M-1}\Bigl\{\mathrm{Re}\Bigl[\mathrm{Var}_{\mathbb C}(\beta)\sum_{n\neq m}\sqrt{w_nw_m}\,e^{i(\omega_n+\omega_m)\tau}\Bigr] - \langle\delta\hat N\rangle\sum_{n\neq m}\sqrt{w_nw_m}\cos\bigl[(\omega_n-\omega_m)\tau\bigr]\Bigr\}\nonumber\\
	&= \frac{2\mathcal E_0^2}{M-1}\Bigl\{\mathrm{Re}\bigl[\mathrm{Var}_{\mathbb C}(\beta)\bigl(\mathcal C_1(\tau)^2-\mathcal C_2(\tau)\bigr)\bigr] - \langle\delta\hat N\rangle\bigl(|\mathcal C_1(\tau)|^2-W\bigr)\Bigr\}.
\end{align}

The resulting contributions in $(\Delta\hat E(\tau))^2_{\rm PERM}=\sum_n(\Delta\hat E_n)^2+\sum_{n\neq m}\mathrm{Cov}(\hat E_n,\hat E_m)$ can then be separated. The vacuum floor is $\mathcal E_0^2W=\mathcal E_0^2M(M+2n_0)$. The terms proportional to $\langle\delta\hat N\rangle$ combine as
\begin{align}
	2\mathcal E_0^2\langle\delta\hat N\rangle\,W - \frac{2\mathcal E_0^2}{M-1}\langle\delta\hat N\rangle\bigl(|\mathcal C_1(\tau)|^2-W\bigr) = \frac{2\mathcal E_0^2}{M-1}\langle\delta\hat N\rangle\Bigl[(M-1)W+W-|\mathcal C_1(\tau)|^2\Bigr] 
	\nonumber\\= \frac{2\mathcal E_0^2}{M-1}\langle\delta\hat N\rangle\Bigl[MW-|\mathcal C_1(\tau)|^2\Bigr],
\end{align}
The terms proportional to $\mathrm{Var}_{\mathbb C}(\beta)$ combine as
\begin{equation}
	-2\mathcal E_0^2\,\mathrm{Re}\bigl[\mathrm{Var}_{\mathbb C}(\beta)\mathcal C_2(\tau)\bigr] + \frac{2\mathcal E_0^2}{M-1}\mathrm{Re}\bigl[\mathrm{Var}_{\mathbb C}(\beta)\bigl(\mathcal C_1(\tau)^2-\mathcal C_2(\tau)\bigr)\bigr] = \frac{2\mathcal E_0^2}{M-1}\,\mathrm{Re}\Bigl[\mathrm{Var}_{\mathbb C}(\beta)\bigl(\mathcal C_1(\tau)^2-M\,\mathcal C_2(\tau)\bigr)\Bigr],
\end{equation}
Collecting all three pieces reproduces equation~\eqref{eq:total-variance-perm}.

\subsection{Special case}\label{specialcase}

For the special case treated in the part, $M\times M$ matrices with $G_{11}=1$, $G_{1v}=\kappa\,(v\geq2)$, $G_{v1}=\kappa^*\,(v\geq2)$, $G_{vw}=1\,(v,w\geq2)$ occur. The structure of $\mathbf K$, together with row-modified descendants of it, can be obtained by expanding along a single row. We use the identities below:
\begin{itemize}
	\item[(i)] If $p$ rows are of constant $c$ value, and the remaining $n-p$ rows are constant $1$: the matrix permanent is $n!\,c^p$. 
	\item[(ii)] If all $n$ rows are identical, $(r_1,\ldots,r_n)$, the permanent is $n!\prod_ir_i$.
	\item[(iii)] The matrix itself has a permanent $(n-1)!\,[1+(n-1)|\kappa|^2]$. 
	\item[(iv)] If row~1 is replaced by $(\lambda,\mu,\ldots,\mu)$, the permanent becomes $(n-1)!\,[\lambda+(n-1)\mu\kappa^*]$. This is easy to see by expanding by the first row.
	\item[(v)] If a regular row $v\geq2$ is replaced by $(\lambda',\mu',\ldots,\mu')$, while row~1 is unmodified, the permanent becomes $(n-1)!\,\{\lambda'\kappa+\mu'[1+(n-2)|\kappa|^2]\}$. Expanding along row~$v$: removing column~1 leaves row~1 (all $\kappa$) and $n-2$ other regular rows (all $1$) [permanent being $(n-1)!\kappa$, by (i)]; removing any other column leaves the configuration of (iii) at size $n-1$ [permanent being $(n-2)![1+(n-2)|\kappa|^2]$].
	\item[(vi)] If there are two identical modified rows $(\lambda',\mu',\ldots,\mu')$ and no unmodified special row are present, with the remaining $n-2$ rows regular, the permanent is $(n-1)!\,\mu'\,[2\lambda'+(n-2)\mu'\kappa^*]$. Expanding along one of the two rows: removing column~1 leaves the other identical row (now constant $\mu'$) plus $n-2$ regular rows [permanent being $(n-1)!\mu'$, by (i)]. Removing any other column leaves the configuration of (iv) at size $n-1$ [permanent being $(n-2)!\,(\lambda'+(n-2)\mu'\kappa^*)$].
	\item[(vii)] If row~1 is replaced by $(\lambda,\mu,\ldots,\mu)$ and simultaneously, one regular row $v$ by $(\lambda',\mu',\ldots,\mu')$, the permanent is $(n-1)!\,\{\lambda\mu'+\mu\lambda'+(n-2)\mu\mu'\kappa^*\}$. Expanding along row~1: removing column~1 leaves row~$v$ (now constant $\mu'$) plus $n-2$ regular rows of permanent $(n-1)!\mu'$; removing any other column leaves the configuration of (iv) at size $n-1$, with $(\lambda',\mu')$ in place of $(\lambda,\mu)$, and permanent $(n-2)!\,(\lambda'+(n-2)\mu'\kappa^*)$.
\end{itemize}
By identity (iii), $\mathrm{perm}(\mathbf K)=(M-1)!\,[1+(M-1)|\kappa|^2]$, and $\mathcal N=M!\,(M-1)!\,[1+(M-1)|\kappa|^2]$.
Direct evaluation of $\mathbb E(\beta^l)=\frac1M\sum_k\beta_k^l$ gives $\mathbb E(\beta^l)=[\alpha^l+(M-1)\beta_0^l]/M$; squaring and subtracting,
\begin{equation}
	\mathbb E(\beta^2)-\mathbb E(\beta)^2 = \frac{M[\alpha^2+(M-1)\beta_0^2]-[\alpha+(M-1)\beta_0]^2}{M^2} = \frac{(M-1)(\alpha-\beta_0)^2}{M^2},
\end{equation}
since $M[\alpha^2+(M-1)\beta_0^2]-[\alpha+(M-1)\beta_0]^2=(M-1)\alpha^2-2(M-1)\alpha\beta_0+(M-1)\beta_0^2$ $=(M-1)(\alpha-\beta_0)^2$, one can write
\begin{align}
	\mathbb E(\beta^l) = \frac{\alpha^l+(M-1)\beta_0^l}{M},
	\qquad
	\mathrm{Var}_{\mathbb C}(\beta) = \frac{M-1}{M^2}(\alpha-\beta_0)^2.  \qquad  \qquad
	\label{eq:special-A-moments}
\end{align}
The mean field is
$\langle\hat{E}(\tau)\rangle_{\mathrm{PERM}}
=
-\frac{2\mathcal{E}_0}{M}\,\bigl|\alpha+(M-1)\beta_0\bigr|
\sum_{n=1}^{M}\sqrt{2n_0+2n-1}\;
\sin\!\bigl(\omega_n\tau+\varphi_0\bigr)$, where $\varphi_0=\arg[\alpha+(M-1)\beta_0]$.

Evaluation of the mean photon number~\eqref{eq:photonmean} requires $\mathrm{perm}(\mathbf K^{[j]}_1)$ for $j=1$ and $j\geq2$. For $j=1$: row~1 becomes $(\beta_1\cdot1,\beta_2\kappa,\ldots,\beta_M\kappa)=(\alpha,\beta_0\kappa,\ldots,\beta_0\kappa)$, identity (iv) with $\lambda=\alpha,\mu=\beta_0\kappa$:
\begin{equation}
	\mathrm{perm}\bigl(\mathbf K^{[1]}_1\bigr) = (M-1)!\,\bigl[\alpha+(M-1)\beta_0|\kappa|^2\bigr].
\end{equation}
For $j\geq2$ (all equivalent by symmetry): row~2 becomes $(\beta_1\kappa^*,\beta_2,\ldots,\beta_M)=(\alpha\kappa^*,\beta_0,\ldots,\beta_0)$. Identity (v) with $\lambda'=\alpha\kappa^*,\mu'=\beta_0$ leads to
\begin{equation}
	\mathrm{perm}\bigl(\mathbf K^{[2]}_1\bigr) = (M-1)!\,\bigl\{\alpha|\kappa|^2+\beta_0[1+(M-2)|\kappa|^2]\bigr\}.
\end{equation} 
Weighting the $j=1$ term by $\alpha^*$ and the $(M-1)$ equivalent $j\geq2$ terms by $\beta_0^*$ in equation~\eqref{eq:photonmean},
\begin{equation}\label{specialcasemeanN}
	\langle\hat N_n\rangle = \frac{|\alpha|^2+(M-1)|\kappa|^2(\alpha^*\beta_0+\alpha\beta_0^*)+(M-1)|\beta_0|^2[1+(M-2)|\kappa|^2]}{M[1+(M-1)|\kappa|^2]}.
\end{equation}
Subtracting $|\mathbb E(\beta)|^2=|\alpha+(M-1)\beta_0|^2/M^2$ from the mean photon number, after some simplification leads to
\begin{equation}
	\langle\delta\hat{N}\rangle
	=
	\langle\hat N_n\rangle-|\langle\hat a_n\rangle|^2
	=
	\frac{(M-1)(1-|\kappa|^2)\,|\alpha-\beta_0|^2}{M^2\bigl[1+(M-1)|\kappa|^2\bigr]}.
	\label{eq:special-A-DeltaN}
\end{equation}

Since $\arg\mathrm{Var}_{\mathbb C}(\beta)=2\arg(\alpha-\beta_0)$, substituting ~\eqref{eq:special-A-moments} and~\eqref{eq:special-A-DeltaN} into equation~\eqref{eq:single-mode-variance-full} gives
\begin{equation}
	\bigl(\Delta\hat E_n\bigr)^2_{\mathrm{PERM}}
	=
	\mathcal E_0^2\,(2n_0+2n-1)
	\Bigl[
	1+2\langle\delta\hat{N}\rangle
	-2\bigl|\mathrm{Var}_\mathbb{C}(\beta)\bigr|\,
	\cos\!\bigl(2\omega_n\tau+2\arg(\alpha-\beta_0)\bigr)
	\Bigr].
	\label{eq:special-A-Evar-single}
\end{equation}
Substituting equations~\eqref{eq:special-A-moments} and~\eqref{eq:special-A-DeltaN} into equation~\eqref{eq:cov-summary}, and using $|\kappa|^2=e^{-|\alpha-\beta_0|^2}$, gives, for $n\neq m$,
\begin{equation}
	\langle\hat a_n\hat a_m\rangle-\langle\hat a_n\rangle\langle\hat a_m\rangle = -\frac{(\alpha-\beta_0)^2}{M^2}\,e^{-i(\omega_n+\omega_m)t}, \qquad
	\langle\hat a_n^\dagger\hat a_m^\dagger\rangle-\langle\hat a_n^\dagger\rangle\langle\hat a_m^\dagger\rangle = -\frac{(\alpha^*-\beta_0^*)^2}{M^2}\,e^{i(\omega_n+\omega_m)t},
	\label{eq:special-A-aaupdown}
\end{equation}
\begin{equation}
	\langle\hat{a}^\dagger_n\hat{a}_m\rangle
	-\langle\hat{a}^\dagger_n\rangle\langle\hat{a}_m\rangle
	=
	-\frac{\bigl[1-e^{-|\alpha-\beta_0|^2}\bigr]\,|\alpha-\beta_0|^2}
	{M^2\bigl[1+(M-1)e^{-|\alpha-\beta_0|^2}\bigr]}\;
	e^{i(\omega_n-\omega_m)t},
	\label{eq:special-A-lambda}
\end{equation}
with $\langle\hat a_n\hat a_m^\dagger\rangle-\langle\hat a_n\rangle\langle\hat a_m^\dagger\rangle$ the complex conjugate of equation~\eqref{eq:special-A-lambda}, since $\langle\delta\hat N\rangle$ is real.
Inserting equations~\eqref{eq:special-A-moments} and~\eqref{eq:special-A-DeltaN} into equation~\eqref{eq:total-variance-perm}, and using $\mathrm{Var}_{\mathbb C}(\beta)/(M-1)=(\alpha-\beta_0)^2/M^2$, gives the total field variance for this configuration,
\begin{align}
	\bigl(\Delta\hat E(\tau)\bigr)^2_{\rm PERM} = \mathcal E_0^2 M(M+2n_0) 
	+ \frac{2\mathcal E_0^2}{M^2}
	\left\{\frac{(1-|\kappa|^2)|\alpha-\beta_0|^2}{1+(M-1)|\kappa|^2}
	\Bigl[M^2(M+2n_0)- \bigl|\mathcal C_1(\tau)\bigr|^2
	\Bigr] \right.
\nonumber\\ 
\left. + \mathrm{Re}\Bigl[(\alpha-\beta_0)^2
	\Bigl(\mathcal C_1(\tau)^2 -M\,\mathcal C_2(\tau) \Bigr)\Bigr]\right\} ,
\end{align}
with $\mathcal C_1(\tau),\mathcal C_2(\tau)$ as in Appendix~\ref{sumvari}. 

Dividing equation~\eqref{eq:special-A-DeltaN} by $|\mathrm{Var}_{\mathbb C}(\beta)|=\tfrac{M-1}{M^2}|\alpha-\beta_0|^2$ gives
\begin{equation}
	\frac{\langle\delta\hat N\rangle}{|\mathrm{Var}_{\mathbb C}(\beta)|} = \frac{1-|\kappa|^2}{1+(M-1)|\kappa|^2} \in [0,1],
\end{equation}
so $\langle\delta\hat N\rangle\leq|\mathrm{Var}_{\mathbb C}(\beta)|$ throughout. By equation~\eqref{eq:special-A-Evar-single}, the single-mode variance attains its minimum, over $\tau$, at the bracket value $1+2\langle\delta\hat N\rangle-2|\mathrm{Var}_{\mathbb C}(\beta)|$; the inequality above gives
\begin{equation}
	1+2\langle\delta\hat N\rangle-2|\mathrm{Var}_{\mathbb C}(\beta)| \;\leq\; 1+2\langle\delta\hat N\rangle - 2\langle\delta\hat N\rangle \;=\;1,
\end{equation}
i.e.\ the single-mode variance dips \emph{to or below} the vacuum-floor bracket value $1$ for every finite $\alpha,\beta_0$.
The magnitude of equation~\eqref{eq:special-A-lambda},
\begin{equation}
	\bigl|\langle\hat a_n^\dagger\hat a_m\rangle-\langle\hat a_n^\dagger\rangle\langle\hat a_m\rangle\bigr| = \frac{\bigl[1-e^{-y}\bigr]\,y}{M^2\bigl[1+(M-1)e^{-y}\bigr]}, \qquad y\equiv|\alpha-\beta_0|^2,
\end{equation}
vanishes at $y=0$ (the coherent-state limit $\alpha=\beta_0$) and grows without bound as $y\to\infty$, approaching $y/M^2$ once $e^{-y}$ is negligible.

One-mode second-order coherence can be evaluated by a lengthy but similar route as $\langle\hat N_n\rangle$ [identities (iv),(v) again, now at power $l=2$: $\lambda\to\alpha^2,\mu\to\beta_0^2\kappa$ for $j=1$; $\lambda'\to\alpha^2\kappa^*,\mu'\to\beta_0^2$ for $j\geq2$], equation~\eqref{adagkal} with $k=l=2$ gives
\begin{equation}
	\langle\hat a_n^{\dagger2}\hat a_n^2\rangle
	=
	\frac{|\alpha|^4+2(M-1)|\kappa|^2\,\mathrm{Re}\bigl[(\alpha^*\beta_0)^2\bigr]
		+(M-1)|\beta_0|^4\bigl[1+(M-2)|\kappa|^2\bigr]}
	{M\bigl[1+(M-1)|\kappa|^2\bigr]},
	\label{eq:special-A-adag2a2}
\end{equation}
and dividing by $\langle\hat N_n\rangle^2$ gives $g^{(2)}_n$.
\\
Two-mode second-order coherence can be evaluated in the following way: $\langle\hat N_n\hat N_m\rangle$ ($n\neq m$) requires $\mathrm{perm}(\mathbf K^{[j,j']}_{1,1})$ from equation~\eqref{twomode} with $k=l=f=g=1$, in two cases according to whether $\{j,j'\}$ includes the special index~$1$.
\\Case A : $\{j,j'\}=\{1,p\}$, $p\geq2$. Row~1 and row~$p$ are simultaneously replaced as for $\mathbf K^{[1]}_1$ and $\mathbf K^{[2]}_1$ above, i.e.\ $\lambda=\alpha,\mu=\beta_0\kappa$ and $\lambda'=\alpha\kappa^*,\mu'=\beta_0$ in identity (vii):
$
\mathrm{perm}\bigl(\mathbf K^{[1,p]}_{1,1}\bigr) = (M-1)!\,\Bigl\{\alpha\beta_0+\beta_0\kappa\cdot\alpha\kappa^*+(M-2)\beta_0\kappa\cdot\beta_0\kappa^*\Bigr\} = (M-1)!\,\beta_0\Bigl[\alpha(1+|\kappa|^2)+(M-2)|\kappa|^2\beta_0\Bigr]$.
\\Case B : $\{j,j'\}=\{p,q\}\subset\{2,\ldots,M\}$. Rows $p,q$ are both replaced by $(\alpha\kappa^*,\beta_0,\ldots,\beta_0)$; row~1 is untouched. Expanding along row~1 [as for identity (v)]: removing column~1 leaves both modified rows constant $\beta_0$ and $M-3$ regular rows constant $1$ [by (i)]. Removing any column $k\geq2$ leaves two identical modified rows among $M-3$ regular rows, i.e.\ identity (vi) at size $M-1$ with $\lambda'=\alpha\kappa^*,\mu'=\beta_0$ [the permanent being $(M-2)!\,\beta_0\kappa^*[2\alpha+(M-3)\beta_0]$]. Hence
$\mathrm{perm}\bigl(\mathbf K^{[p,q]}_{1,1}\bigr) = (M-1)!\,\beta_0^2 + (M-1)\kappa\,(M-2)!\,\beta_0\kappa^*\bigl[2\alpha+(M-3)\beta_0\bigr] = (M-1)!\,\beta_0\Bigl[\beta_0\bigl(1+(M-3)|\kappa|^2\bigr)+2\alpha|\kappa|^2\Bigr]$.

Of the $M(M-1)$ ordered pairs $(j,j')$ in equation~\eqref{twomode}, $2(M-1)$ have $\{j,j'\}=\{1,p\}$ (weighted by $\alpha^*\beta_0^*$; both orderings give the same permanent, since $\mathbf K^{[j,j']}_{1,1}=\mathbf K^{[j',j]}_{1,1}$ when $l=g$) and $(M-1)(M-2)$ have $\{j,j'\}\subset\{2,\ldots,M\}$ (weighted by $\beta_0^{*2}$). So
\begin{equation}
	\langle\hat N_n\hat N_m\rangle = \frac{2(M-1)\alpha^*\beta_0^*\,\mathrm{perm}\bigl(\mathbf K^{[1,p]}_{1,1}\bigr) + (M-1)(M-2)\beta_0^{*2}\,\mathrm{perm}\bigl(\mathbf K^{[p,q]}_{1,1}\bigr)}{M(M-1)\,\mathrm{perm}(\mathbf K)},
\end{equation}
which, collecting terms into real combinations by the same route as for $\langle\delta\hat N\rangle$ above, reduces to
\begin{equation}
	\langle\hat N_n\hat N_m\rangle
	=
	\frac{|\beta_0|^2\Bigl\{
		2|\alpha|^2(1+|\kappa|^2)+4(M-2)\,\mathrm{Re}(\alpha^*\beta_0)\,|\kappa|^2
		+(M-2)|\beta_0|^2\bigl[1+(M-3)|\kappa|^2\bigr]
		\Bigr\}}
	{M\bigl[1+(M-1)|\kappa|^2\bigr]}.
	\label{eq:special-A-NnNm}
\end{equation}
Dividing by $\langle\hat N_n\rangle\langle\hat N_m\rangle=\langle\hat N_n\rangle^2$ gives the two-mode second-order cross-coherence $g^{(2)}_{nm}$.

\appendix

\bibliographystyle{iopart-num-long}
\bibliography{allthebibGA}

\end{document}